\documentclass[11pt]{article} 
\usepackage[utf8]{inputenc} 
\usepackage{geometry} % to change the page dimensions
\usepackage{graphicx} % support the \includegraphics command and options

\usepackage{booktabs} % for much better looking tables
\usepackage{array} % for better arrays (eg matrices) in maths
\usepackage{paralist} % very flexible & customisable lists (eg. enumerate/itemize, etc.)
\usepackage{verbatim} % adds environment for commenting out blocks of text & for better verbatim
\usepackage{subfig} % make it possible to include more than one captioned figure/table in a single float
\usepackage{changes}
\usepackage{bbm}
\usepackage{amsmath,amsfonts,amssymb}
\usepackage{ulem}
\usepackage[active]{srcltx}
\usepackage{hyperref}
\hypersetup{colorlinks=true,pdfborder={0 0 0}}
\usepackage{fancyhdr} % This should be set AFTER setting up the page geometry
\usepackage{sectsty}
\allsectionsfont{\sffamily\mdseries\upshape} % (See the fntguide.pdf for font help)
\usepackage[nottoc,notlof,notlot]{tocbibind} % Put the bibliography in the ToC
\usepackage[titles,subfigure]{tocloft} % Alter the style of the Table of Contents

\def \be{\begin{equation}}
\def \ee{\end{equation}}
\def\id{\mathbbm{1}}

\newcommand{\bra}[1]{\left\langle #1 \right|}
\newcommand{\braket}[2]{\langle #1 | #2 \rangle}
\newcommand{\ket}[1]{\left| #1 \right\rangle}
\newcommand{\der}[2]{\frac{d #1}{d #2 }}
\newcommand{\e}{\varepsilon}

\newcommand{\pder}[2]{\frac{\partial #1}{\partial #2 }}
\newcommand{\Ran}{\operatorname{\mathrm{Range}}}
\newcommand{\mbf}[1]{\mathbf{ #1} }
\newcommand{\green}[1]{{\color{green} #1}}
\numberwithin{equation}{section}
\newtheorem{exa}{Example}[section]
\newtheorem{rem}{Remark}[section]
\newtheorem{claim}{Claim}[section]
\title{Braiding fluxes  in Pauli Hamiltonian}

\author{O. Kenneth and J.E. Avron\\
{\small  Department of Physics, Technion, 3200003 Haifa, Israel} 
\\
\footnotesize {\texttt{ email: kenneth@physics.technion.ac.il}}}

\begin{document}
\maketitle
\normalem
	\begin{abstract}
 
Aharonov and Casher  showed that Pauli Hamiltonians in two dimensions  have gapless zero modes. 
We study the adiabatic evolution of these modes under the slow motion of $N$ fluxons with fluxes $\Phi_a\in\mathbb{R}$.  
The positions, $\mbf{r}_a\in\mathbb{R}^2$, of the fluxons are viewed as controls. We are interested in the holonomies 
associated with closed paths in the space of controls.  The holonomies can sometimes be abelian, but in general are not. 
They can sometimes be topological, but in general are not. We analyze some of the special cases and some of the general ones. 
Our most interesting results concern the cases where  holonomy turns out to be topological which 
is the case when all the fluxons are  subcritical, $\Phi_a<1$,  and the 
number of zero modes is $D=N-1$.  If $N\ge3$ it is also non-abelian.  In the special case that the fluxons carry  identical  
fluxes the resulting anyons satisfy the Burau representations of the braid group.     

	\end{abstract}
%%%%%%%%%%%%%%%%%%%%%%%%%%%%%%%%%%%%%%%%%%
	\section{Introduction}

The Pauli Hamiltonian describes a non-relativistic electron with gyromagnetic constant $g=2$
\begin{align}\label{Pauli2}
	\mbf{H}_p(A)
	&=\frac 1{ 2m} 
	\left(-i\nabla-e\mathbf{A}\right)^2\otimes \id - \frac {ge}
	{4m} \mathbf{B}\cdot\sigma -eA_0 \otimes \id\, 
	\end{align}
$\sigma$ is the vector of Pauli matrices and $\mbf{H}_p$  acts on spinors. { We use  units where $\hbar=c=1$.} 
The  electric and magnetic fields are  {determined by the 4-potential $A=(A_0,\mathbf {A})$}:
	\be \label{fields}
	\mathbf{B}=\nabla\times\mathbf{A},\quad \mathbf{E}=-\pder{}{t} 
	\mathbf{A}+\nabla A_0
	\ee

In 1979 Aharonov and Casher \cite{AharonovCasher} observed that the Pauli operator  for static magnetic field in two dimensions, 
$ {A}=(0,A_x,A_y,0)$, so $ \mathbf{B}= B\,\mathbf{\hat z}$, has  (normalizable) zero energy modes
\footnote{When  $g>2$ the zero modes turn into gapped bound states.}.  
They are gapless ground states and their number $D$, is determined by  the total magnetic flux $\Phi_T$ measured in units of 
quantum flux, 
\be\label{zero-modes}
	D= \left\lceil|\Phi_T|\right\rceil-1
       , \quad \Phi_T=\frac{e}{2\pi} \int {B}\,  dx\wedge dy%,\quad  \Phi_0=\frac{2\pi}e  
	\ee
%%%%%%%%%%%%%%%%%	
Where $\lceil x\rceil$ stands for the Ceiling of $x$, i.e. the smallest integer $\geq x$.
%%%%%%%%%%%%%%%%

	 \begin{figure}[htb]
    	\centering
   	\includegraphics[width=6 cm]{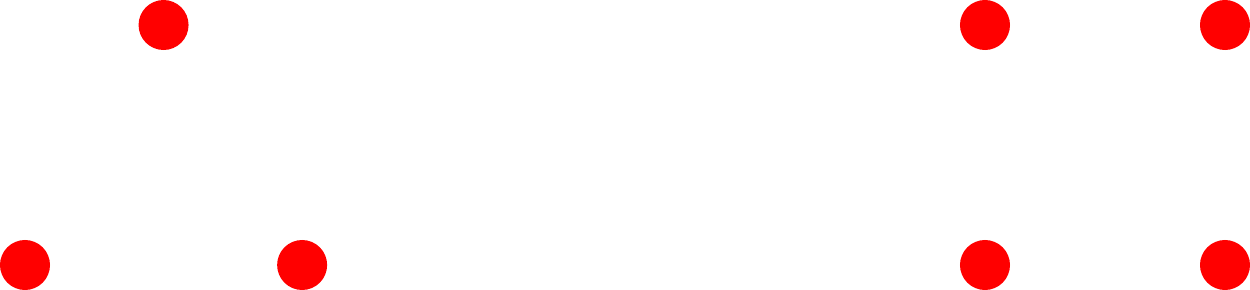}
\medskip
	\caption{The figure shows two clusters of isolated fluxons. If each fluxon is  critical, $\Phi_a=1$,   
the total number of zero modes is 6.  
Since all the fluxons are critical no mode is localized on any one fluxon.  If the two clusters are well 
separated one can choose two of the 
zero mode to be localized on the cluster of three fluxons and another three modes to be localized on the cluster 
of four fluxons. The remaining 6-th mode  is supported on the two clusters no matter how far they are separated. }\label{fig:qubits}
	\end{figure}

We consider a magnetic field $B$ localized on a finite number of disjoint fluxons labeled by $a=1,\dots, N$.  
The magnetic flux of the $a$-th fluxon, $\Phi_a$,  is  localized in a {region} of radius $R_a$ centered at $\mathbf{r}_a$. 
  We {\it do not}  assume that $\Phi_a$ is quantized or that all the fluxes 
$\Phi_a$ are identical.  
{We shall}  {assume w.l.o.g. that $\Phi_T>0$.}
We say that the $a$-th fluxon is super-critical if $\Phi_a>1$, subcritical if $\Phi_a<1$ and critical 
if   $\Phi_a=1$. The fluxons are viewed as classical parameters and {\it not} as dynamical degrees of freedom: They  {\it do not} 
have a wave function or an equation of motion\footnote{In contrast with,  \cite{AharonovColeman}, where the fluxons have 
a wave-function and an equation of motion, and are assumed to carry half a unit of quantum flux.}. 
(The dynamical degree of freedom is the electron wave function.)

%%%%%%%%%%%%%%%%%%%%%%%%%%%%%%%%%0

%%%%%%%%%%%%%%%%%%%%%%%%%%%%%%%%1

%\red{ 
When the $a$-th fluxon is super-critical  it can create $D_a=\lceil\Phi_a\rceil-1\neq0$
zero modes which are {\it confined} to it, in the sense that their wave function decays
(as a power law) over a typical distance $O(R_a)$ determined by the fluxon radius $R_a$.
More interesting  are the zero modes which are bound jointly by a number of separate fluxons.
We shall call these solutions {\it free zero modes}.
These states wave functions live in between the fluxons and have typical size determined
by the inter-distance  $|\mathbf{r}_a-\mathbf{r}_b|$.
When {\it all} the fluxons are subcritical, $D_a=0$ {\it all} the zero modes are free: 
The probability of finding the charge on any of the fluxons  is close to zero  (as $R_a\to 0$).  
In general, confined and free modes coexist.  
The confined modes behave like the charge-flux composites one encounters in the fractional quantum Hall effect 
\cite{jain,preskill,arovas,wilczek}, except that here the  charge is quantized but the flux is not whereas in the Hall 
effect it is the flux that is quantized and the charge is not. 
The free modes are a different kettle of fish as the composite involves a single electron jointly bound by several fluxons. 
As we shall see, these modes can sometimes turn the fluxons into non-abelian anyons  \cite{kitaev2003fault,preskill,frohlich}. 
These new ``topological''  objects are quite  interesting.

%%%%%%%%%%%%%%%%%%%%%%%%%%%2

 The distinction between confined and free  zero modes is meaningful when 
the radius of the {individual} fluxons, $R_a$, is the smallest length scale in the problem, 
$R_a\ll |\mathbf{r}_a-\mathbf{r}_b|$  and is sharp  for point-like fluxons.  
The total number of free  modes {$D_f$} is, as we shall see,
%%   \be\label{extended}
%%   0\le D_f= \left\lceil \left|\sum_a \Phi_a'\right|\right\rceil -1\le N-1, 
%%   \quad     n_a=Max\{ \lfloor \Phi_a\rfloor,0\},\, \Phi'_a=\Phi_a-n_a 
%%   \ee
{   \be\label{extended}
0\le D_f= Max\{0,\left\lceil\sum_a \Phi_a'\right\rceil -1\}\le N-1, \quad    \Phi'_a=\Phi_a-D_a     \ee    }
%%%%%%%%%%
We say that the number of free modes is maximal if  $D_f=N-1$. This turns out to be the 
case where the fluxons become non-abelian anyons.
If all the fluxons are identical then to have maximal number of free modes leading to interesting
representation of the braid group requires
\be\label{con}
1-\frac 1 N <\Phi_a< 1
\ee
(The case $\Phi_a\equiv1$ leads to a trivial representation of the pure braid group.)

To study the holonomy of the zero modes we treat the fluxon coordinates, 
$\mathbf{r}_a$ as  (classical) adiabatic controls.  
The adiabatic theory we shall need and describe is of interest in its own right, since the weak 
electric fields generated by the slow motion of the fluxons are important for the adiabatic transport 
and since the zero modes are gapless  (see Sec.~\ref{sec:ad} for more details).  
%%%%%%%%%%HHHHHHHHHHHHHH%%%%%%%%%%%%%%
{Adiabaticity means  that the characteristic time scale of the controls is large compared with 
the characteristic time scale of the system. We shall argue that the characteristic time scale in the  case of point-like fluxons 
is set by their mutual distances.}
This means that points in control space where fluxons collide must be removed: 
Fluxon collisions is like gap closures in gapped systems. Both endow  control space with an interesting topology 
which is sine qua non for interesting topological behaviour.

\subsection{Holonomies:}

The holonomies of braiding point-like\footnote{We shall not be concerned with a rigorous study of the limit of point-like 
fluxons aka point interactions \cite{exner}.} fluxes are summarized in:

 	\begin{itemize}
 	\item 
The Berry phase  associated with the confined mode on the $a$-th super-critical fluxon braided by the fluxon $\Phi_b$  
is  the Aharonov-Bohm phase $2\pi \Phi_b$.  

%%%   ***************************
%%%
%%%   \item  When the number of free modes is maximal $D_f=N-1$ the holonomy is {\it topological} yet not trivial. 
%%%   When the number of free zero modes is smaller $D_f<N-1$, the holonomy is 
%%%   path {\it dependent}. The adiabatic curvature is then non zero, see Fig. \ref{fig:curvature}.
%%%
%%%
%%% @@@@@@@@@@@@@@@@@@

	\item  The Berry phase for a non-degenerate free mode, ($D_f=1$), and two fluxons ($N=2$) is   
{\em topological} (path independent) given by $2\pi(\Phi_T-1)$. 
	\item  The Berry phase for a single free mode, ($D_f=1$) and $N\ge 3$ fluxons is 
abelian but path {\it dependent}. In other words, the adiabatic curvature is non trivial, see Fig. \ref{fig:curvature}.
	\item  For $N\geq 3$ and maximal number of free modes, 
$D_f=N-1$ the holonomy is non-abelian and
{{\it topological}}.  Braiding  anyon $a$ with anyon $b$ is associated with the monodromy matrix
\be
{\left(\begin{array}{cc}
1-\nu_a+\nu_a\nu_b & \nu_a(1-\nu_b) 
\\
1-\nu_a  & \nu_a
  \end{array}\right )},\quad  \nu_a=e^{-2\pi i\Phi_a}
 \ee
 The eigenvalues of the holonomy matrix are $\{1, \nu_a\nu_b\}$. This is our most significant result.
 \item  If, in addition to $D_f=N-1$, all the fluxons carry identical fluxes, 
then exchanging them make sense and  
is described by  matrices that give  the Burau representation 
of the full braid group \cite{birman}:
\be
{\left(\begin{array}{cc} 
1-\nu_a & \nu_a\\ 
 1 & 0
\end{array}\right )}, \quad 
\ee
 The fluxons are  { identical} anyons.    
Like ordinary anyons \cite{kitaev2003fault, preskill}  they have topological braiding  and fusion rules. 
(The fusion rules are simply flux addition.)  But, unlike ordinary anyons, they are gapless and hence fragile.  
	\item $N\ge 3$ and $1<D_f<N-1$: The holonomy is non-abelian and, in general, path dependent {i.e. not topological}.
  	\end{itemize}

%%%%%%%%%%%%%%%%%%%%%%%%%%%%%%%%%%%%%%%
	\section{The Aharonov-Casher Zero modes}\label{sec:AC}

A key to Aharonov-Casher (AC) observation  is the fact that when $A_0=0$ and $g=2$ the Pauli Hamiltonian is a prefect square 
	\begin{align}\label{Pauli3}
	\mbf{H}_p(0,\mathbf{A})
	%&=\frac 1{ 2m} 
		%\left(-i\mathbf{\nabla-\mathbf{A}}\right)^2\otimes \id - \frac {1}
	%{2m} \sigma\cdot \mathbf{B} \nonumber \\
	&=\frac 1{ 2m} 
	\big(	\left(-i\nabla-e\mathbf{A}\right)\cdot\sigma\big)^2
	\end{align}
Since  $H_p(0,\mathbf{A})\ge 0$ the zero modes, if any, are ground states and are the normalizable solutions of 
	\be
	\left(-i\nabla-e\mathbf{A}\right)\cdot\sigma\,\psi=0
	\ee
where $\psi$ is a two component spinor. 

The second key observation  is {special to} two dimensions.   {It is convenient then to use complex notation}
\footnote{Note that $\bar A_z=A_{\bar z}$ and $A_x dx+ A_y dy =A_z dz+ A_{\bar z} d\bar z$}
	\be
	z=x+iy,\;\;\;2\partial_z=\partial_x
	-i\partial_ y,\;\;\; 
	2A_z=A_x-iA_y,\;\; % \mathbf{B}= B\, \mathbf{\hat z}
	\ee
One then has
	\be
	\left(-i\nabla-e\mathbf{A}\right)\cdot\sigma\,%\psi
	=
	-2i \left(\begin{array}{cc} 0 & \partial_z-ie{A_z} \\  
	\bar\partial_z-ie\bar A_z &  0 \end{array}\right)
	%\left(\begin{array}{c} \psi_\uparrow \\  \psi_\downarrow\end{array} \right )=0
	=Q+Q^*
	\ee
{ where $Q=2\sigma_+(-i\partial_z-eA_z)$.}
Since $Q^2=0$ the Pauli Hamiltonian is super-symmetric \cite{witten1982supersymmetry,cycon1987schrodinger}:
	\begin{align}\label{Pauli4}
	H_p(\mathbf{A},0)
	%	&=\frac 1{ 2m} 
	%\big(	\left(-i\mathbf{\nabla-\mathbf{A}}\right)\cdot\sigma\big)^2\\
	&= \frac 1 {2m} \{Q,Q^*\},  \nonumber
	\end{align}
%%%%%%%%%%%%%%%%%%%
The zero-modes then lie in  $\ker Q\cap \ker Q^*$, i.e. they are either spin up states that lie in the kernel 
of $\bar\partial_z -ie\bar A_z$, or spin down states in the kernel of $\partial_z -ie A_z$.
For the zero-modes with spin up:
	\be\label{kernel1}
 	\psi= 
	\left(\begin{array}{c} \psi_\uparrow \\  0\end{array} \right ), 
	\quad (\bar\partial_z-ie\bar A_z)\psi_\uparrow=0
	\ee
Let us first look for a solution  that does not vanish anywhere, so  $\log \psi_\uparrow$ is well defined.  We shall call this 
a fundamental solution and denote it by $\psi_0$ .  It is given by
	\be\label{alpha}
	\bar\partial_z \log\psi_0=ie\bar A_z 
        \Longrightarrow \partial_z\bar\partial_z \log\psi_0=ie\partial_z \bar A_z
  	\ee
  	Using 
  	\be
  	4\partial_z\bar\partial_z=\Delta, \quad 4\partial_z\bar A_z
	=\texttt{div} \mathbf{A}+i \texttt{curl} \mathbf{A}
  	\ee
it follows that $\log\psi_0$ is a solution of Poisson's equation whose source term is determined by $\mathbf{A}$
	\be \label{poisson}
	 \Delta \log\psi_0 = -eB +ie \nabla\cdot\mathbf{A} 
	\ee 
Consequently,  a unique choice of $\log\psi_0$ is made by means of the Poisson kernel: 
	\be
	\Delta^{-1}(z,z')= \frac 1 {2\pi} \log |z-z'|
	\ee
By elliptic regularity $\log\psi_0$ is at least as regular as   $ -B +i \nabla\cdot\mathbf{A} $. 

%%%%%%%%%%%%%%%%%%%
 In the Coulomb gauge  $\nabla\cdot \mathbf{A}=0$. Consequently $\log\psi_0$ is real and the fundamental solution is positive.  Clearly
	\be\label{alphaPK}
	\log\psi_0\underset{z\to \infty}{\longrightarrow} 
	-\Phi_T\,\log|z|
	\ee
with $\Phi_T$ the total magnetic flux. 
%%%%%%%%%%%%%%%
Since $|\psi_0|$ is gauge invariant the fundamental solution (in any gauge) decays polynomially:
	\be
|	\psi_{0}| \underset{z\to \infty}{\longrightarrow} |z|^{-\Phi_T}
	\ee
 The fundamental solution is square integrable  iff 
$\Phi_T>1$ .
 
 Similarly, the spin down fundamental solution decays at infinity if $\Phi_T<0$ and is square integrable iff $\Phi_T<-1$.  
We shall assume from now on that $\Phi_T>0$ and consider only spin up zero modes. 
  
 Now with $P(z)$ any holomorphic, 
 	\be\label{yr7}
 	\psi_\uparrow=P(z)\psi_0\Longrightarrow 
	\psi_\uparrow \in Ker(\bar\partial_z-ie\bar A_z)
 	\ee
$P(z)$ cannot have poles, since this conflicts with (local) square integrability of $\psi_\uparrow$ (and the regularity of $\log\psi_0$).  
$P(z)$ must therefore be a polynomial.  $\psi_\uparrow$ is  square integrable  provided
\[
\deg(P) <\Phi_T-1\]
 It follows that there are  $D$	
zero modes  with $D$ given by Eq.~(\ref{zero-modes}).

These results of Aharonov and Casher (AC) \cite{AharonovCasher} may be viewed as an example of an index theorem for 
non-compact manifold  \cite{cycon1987schrodinger}.
%%%%%%%%%%%%%%%%%%%%%%

	\subsection{Zero modes of fluxons}
%%%%%%%%%%%%%%%%%%%%%%%%%%%%%%%%%%%%%%

We now turn to the study of static disjoint fluxons   with fluxes $\Phi_a$ localized inside discs of radius $R_a$ 
centered at $\mathbf{r}_a$.  We shall denote $\Phi=(\Phi_1,\dots,\Phi_N)\in\mathbb{R}^N$ and $\Phi_T=\sum \Phi_a$.
An interesting and very useful feature of the AC zero modes, which follows from Eq.~(\ref{poisson}), is the `superposition' property
\footnote{Note however that the normalization of $\psi_0$ has no simple relation to the normalization of the factors $(\psi_a)_0$.}:  
The fundamental solution for $N$ fluxons is the product of the single fluxon fundamental solutions:
	\be\label{superposition}
	\psi_0=\prod_a (\psi_a)_0
	\ee
 In particular, in the Coulomb gauge,
\[
 (\psi_a )_0= e^{-e\Delta^{-1} B_a}
\]

%%%%%%%%%%%%%%%%%%%%%%%%%%%%%%%%%%%%

	\subsubsection{ A single fluxon}\label{sf}
Consider  a single fluxon  with uniform $B$ localized in a disc of radius $R$ centered at the origin. In the Coulomb gauge 
	\be\label{Buniform}
eB(\mathbf{r})=\frac{2\Phi_T}{ R^2}\left\{\begin{array}{lr} 
	1 & r<R \\  
	0   & r>R 
	\end{array}\right. 
	\Longleftrightarrow
 	e\mathbf {A}=\hat\theta\,\frac{\Phi_T}{ r}
 	\begin{cases}
 	\left(\frac{r}{R}\right)^2& r<R\\  1 & r>R
 	\end{cases}
 	\ee
The fundamental solution (in the Coulomb gauge) is,  by  Eq.~(\ref{poisson}),  
\be\label{Rnz}
	\psi_0=
	\left\{\begin{array}{lr} \exp\left(-\frac{ r^2}{2 R^2}\Phi_T\right) & r<R \medskip  \\ 
              \left(\frac R  r\right)^{\Phi_T}  \exp\left({-\frac{1} {2}\Phi_T}\right) & r>R 
	\end{array}\right.
	\ee

For $\Phi_T\le 1$ the fundamental solution is not square integrable near infinity.  A sub-critical fluxon can not support zero modes.  
When the total flux is super-critical $\Phi_T>1$ the fundamental solution is square integrable for any $R>0$
and it is spread over an area of typical diameter $O(R)$. 
Similar conclusions hold for  {asymmetric }  $B(\mathbf{r})$ with the same total flux $\Phi_T$ since the asymptotics
of Eq.~(\ref{Rnz}) remains unchanged.

%%%%%%%%%%%%%%%%%%%%0

%%%%%%%%%%%%%%%%%%%%1

It is instructive to compare this result with the formal solution for  single point fluxon.
The fundamental solution for a delta localized magnetic field  is
	 \be\label{point}
	\psi_0= r^{-\Phi_T}, 
	 \ee
which is {\it never} square integrable, in contrast with what we found for finite $R$; The limit $R\to 0$ must be taken with care.
%%%%%%%%%%%%%%%%%
%\red{
In our discussion of point-like fluxons we shall assume that $R$ is smaller than any other length scale in our system
but is actually non-zero and can therefore serve as a small distance cutoff.
This issue will however not be relevant if all fluxons are sub-critical, in which case one may safely put $R=0$.
%} 

%%%%%%%%%%%%%%%%%%%%%%%%%%%%%%%%%%%%%%%%%%%
	\subsubsection{The zero modes of sub-critical point-like fluxons}\label{sec:extended}

Consider $N$  sub-critical, point-like  fluxons.
By the superposition property the fundamental solution is
just a product of (translated) solutions of the form Eq.~(\ref{point}):
	\be\label{prod}
	\psi_0(\mathbf{r})=\prod|\mathbf{r}-\mathbf{r}_a|^{-\Phi_a}
	\ee
The solution is locally square integrable since, by assumption,  the individual fluxons are sub-critical $\Phi_a<1$ and distinct 
$\mathbf{r}_a\neq \mathbf{r}_b$.  It is square integrable at infinity if the total flux is super-critical, $\Phi_T\equiv\sum \Phi_a >1$.  
Eq.(\ref{prod}) is the prototype of free zero modes.

When the $a$-th fluxon  collides with the $b$-th fluxons the norm of the fundamental solution  $\braket{\psi_0}{\psi_0} $ diverges 
if $\Phi_a+\Phi_b\ge 1$.   
The condition  $ \braket{\psi_0}{\psi_0}<\infty$  endows the control space of point-like fluxons with a non-trivial topology.

For point fluxes, a useful gauge, besides the Coulomb gauge, is the {\it holomorphic gauge} which formally has  $\mathbf{A}=0$ 
except for cuts where it is delta like. The fundamental solution in this gauge  is a holomorphic function in the cut {complex} plane 
\be\label{holomorphic}
\psi_0(z;\zeta)=\prod_a(z-\zeta_a)^{-\Phi_a},\quad \zeta_a= \mathbf{r}_a\cdot (\mathbf{\hat x} +i\mathbf{\hat y}) 
,\quad \zeta\in\mathbb{C}^N
\ee
$ \psi_{0} (z;\zeta)$ is analytic in 
$\mathbb{C}\backslash\cup\Sigma_a$. The cut $\Sigma_a$  runs from $\zeta_a$ to infinity
\footnote{When the fluxes can be grouped into clusters with total integer flux, the cuts could be organized  in clusters. 
$\psi_0(z;\zeta)$ is then well defined at infinity. This can be understood as reflecting the Dirac flux quantization on compact manifolds.}. 
The general solution obtained by multiplying by a polynomial $P(z)$ is then also holomorphic with the same cuts.
\be\label{lijs}
\psi(z;\zeta)=P(z)\prod_a(z-\zeta_a)^{-\Phi_a},\quad \deg(P) <\Phi_T-1 \ee
%%%%%%%%%%%%%%
In general one may allow having both positive and negative fluxons.

\subsubsection{Free and confined states}\label{super}

%%   In general one may allows having both positive and negative fluxons. For the sake of simplicity we shall assume
%%   \footnote{It is suffice to assume $\Phi_T'>0$.}
%%   in this section that all fluxons have the same sign, $\Phi_a>0$ for all $a$. 

The solution Eq.~(\ref{lijs}) corresponding to $R\rightarrow0$ is typically not square integrable if some of
the fluxons are super-critical.
It becomes a legitimate solution only for  $P(z)\mapsto {P}(z)\prod(z-\zeta_a)^{n_a}$ where $n_a=Max\{0,{\lfloor\Phi_a\rfloor}\}$. 
The corresponding modes are then identical to those occurring in a system of (sub-critical) fluxons having 
$\Phi_a'=\Phi_a-n_a$ (assuming that $\Phi_T'\equiv\sum\Phi_a'>0$).
We call these states free zero modes.
As we shall see their behaviour and holonomies are  identical to those of a system with fluxes $\Phi_a'$.

Since in general $\sum \Phi_a>\sum \Phi_a'$, the system will also contain another type of zero modes. 
In these states the electron typically sits in a small $O(R)$ neighbourhood of a specific fluxon.
We therefore call these states confined. 
In the special case of integer fluxons  certain states may incorporate feature of both confined and free states
(see e.g. Fig. \ref{fig:qubits}).

The main focus of this paper is on the holonomies of free states and the reader may assume, for simplicity,
that all fluxons are subcritical  so that $\Phi_a'=\Phi_a$ and all states are free.
For completeness we give below a brief account of the confined states.

Confined states localized near the $a$'th fluxon are typically constructed by taking
$P(z)\mapsto {P}(z)\prod_{b\neq a}(z-\zeta_b)^{n_b}$ in Eq.~(\ref{yr7}). 
It is convenient to take advantage of our superposition principle and write $\psi$ as a product of 
(not necessarily fundamental) solutions $\psi_b$ corresponding to all fluxons.
A confined state at $a$ then takes the form
\be\psi(z)=\psi_a^j(z)\prod_{b\neq a}\psi_b(z)\simeq\psi_a^j(z)\prod_{b\neq a}\psi_b(\zeta_a).\label{uigf}\ee
Here $\psi_a^j(z),j=0,1..n_a-1$ are some wave-functions  which depend on the detailed shape and radius $R_a$ of the confining 
fluxon, while for $b\neq a$ we have  $\psi_b=(z-\zeta_b)^{n_b}(\psi_b)_0\simeq(z-\zeta_b)^{-\Phi_b'}$ in holomorphic gauge.
The approximate equality on the right of Eq.(\ref{uigf}) follows from the fact that $\psi_a^j(z)$ is sharply peaked near $z=\zeta_a$.

The cases when $\Phi_T'<0$ or $\Phi_a\in\mathbb{Z}$ require a more careful analysis.
We shall not delve into this here since our main interest is in the free modes.

Modes localized on different fluxons are clearly mutually orthogonal.
Some extra thought also shows that when properly normalized the overlap of   free state with a confined state
localized at $\mathbf{r}_a$ scales as $R_a^{1-\Phi_a'}$ and hence vanishes in the pointlike limit.

%%%%%%%%%%%%%%%%%%%%%%%%%%%%%%%%%%%%%%%%%2

%%%%%%%%%%%%%%%%%%%%%%%%%%%%%%%%%%%%%%%%%%%%%%
%%%%%%%%%%%%%%%%%%%%%%%%%%%%%%%%%%%%%%%%%%%%%%
\section{Adiabatic evolution}\label{ch:adiabatic}
We are interested in the evolution of the zero modes
when the fluxes move {adiabatically}. Control space, parametrized by the fluxon coordinates 
$(\mathbf{r}_1,\mathbf{r}_2,...\mathbf{r}_N)\in\mathbb{R}^{2N}$ 
is  $2N$ dimensional. Since the motion of the fluxons generates electric fields we need first to construct corresponding 
(time-dependent)  Pauli operator, Eq.~(\ref{Pauli2}), with both $A_0\neq 0$ and $\mbf{A}\neq0$.

%%%%%%%%%%%%%%%%%%%%%%%%%%%%%%%%
\subsection{The gauge field of moving fluxons}\label{7tuygd}
%%%%%%%%%%%%%%%%%%%%%%%%%%%%%%%

By Faraday law a moving magnetic field must be 
accompanied by a nonzero electric field.
If the motion is  adiabatic the velocity $\mathbf{v}$ is small and the acceleration negligible. It follows that radiation and retardation 
can be neglected.  The fields resulting from the motion can be obtained by Lorentz transformation to the moving frame 
\be \label{evb}
\mathbf{E}=-\mathbf{v}\times\mathbf{B}
\ee 
We therefore need {first to determine} the full Pauli operator,  allowing for both scalar and vector potentials, 
Eq.~(\ref{Pauli2}), due to the motion of fluxons.  The main result  of this subsection is Eq.~(\ref{phi}) which we shall now derive.

%	\be\label{Pauli2}
%	H_p(A)=\frac 1{ 2m} 
%	\big(\left(-i\mathbf{\nabla-\mathbf{A}}\right)\cdot\sigma\big)^2-A_0%, \quad \mathbf {V}=-i\nabla-\mathbf{A}
%	\ee
%where  $\mbf{E}$ and    $\mbf{B}$ are related to the gauge fields by  Eq.~(\ref{fields}). 
To determine $A$   associated with a moving fluxon we substitute  Eq.~(\ref{evb}) in {the definitions of the potentials}, Eq.~(\ref{fields}), 
	\be
	\pder{\mathbf{A}}{t}-\mathbf{\nabla}A_0=-\mathbf{E}=\mathbf{v}\times\mathbf{B}=
	\mathbf{v}\times(\mathbf{\nabla}\times\mathbf{A})=\mathbf{\nabla}		
(\mathbf{v}\cdot\mathbf{A})-(\mathbf{v}\cdot\mathbf{\nabla})\mathbf{A}
	\ee
(and on the right we used the fact that $\mbf v$ is a vector, not a vector field\footnote{
We assume that the fluxons motion is rotation free. Self rotation are expected to affect only the localized modes.
see Appendix~\ref{sec:d} for a discussion of the general case.}.) This may be rearranged as
	\be\label{potentials}
	\der{\mathbf{A}}{t}=\left(\pder{}{t}+\mathbf{v}\cdot\mathbf{\nabla}\right)\mathbf{A}=\mathbf{\nabla}	
(\mathbf{v}\cdot\mathbf{A}+A_0)
	\ee
Let the static magnetic field of the $a$-th fluxon be described by the  
vector potential $\mathbf{A}_a(\mathbf{r})$.  
Take $\mbf{A}$ to be the rigid transport of $\mbf{A}_a$, so that  $\der{\mathbf{A}}{t}=0$, and choose $A_0$ so that  
Eq.~(\ref{potentials}) is satisfied,\footnote{One may check that this is consistent with the sources
$\mathbf{j}(\mathbf{r},t)=\mathbf{j}_a(\mathbf{r}-\mathbf{r}_a(t)),\rho=\mathbf{v}_a\cdot\mathbf{j}$.} namely
%%%%%%%%%%%%%%%%%%   $\vec{j}(\vec{r},t)=\vec{j}_a(\vec{r}-\vec{r}_a(t)),\rho=\vec{v}_a\cdot\vec{j}$.} namely
	\be\label{rot}
	\mathbf{A}(\mathbf{r},t)=\mathbf{A}_a(\mathbf{r}-\mathbf{r}_a(t)),\quad A_0(\mathbf{r},t)=-\mathbf{\dot r}_a\cdot 
	\mathbf{A}_a(\mathbf{r}-\mathbf{r}_a(t))
	\ee
 $\mathbf{r}_a(t)$ is the trajectory of the fluxon.  This 4-potential generates the fields of a rigidly moving fluxon:
	\be\label{fields2}
	\mathbf{B}(\mathbf{r},t)=\mathbf{B}_a(\mathbf{r}-\mathbf{r}_a(t)),\quad\mathbf{E}	
              (\mathbf{r},t)=-\mathbf{\dot r}_a\times \mathbf{B}(\mathbf{r},t)
	\ee
The generalization to a number of fluxons each moving along its own trajectory is obviously
\footnote{Alternatively Eq.~(\ref{phi}) may be derived by applying a Lorentz boost to
the vector $\mathbf{A}_a$ and keeping terms only up to first order in $\mathbf{v}_a$. }
	\be\label{phi}
	\mathbf{A}=\sum_{a=1}^N\mathbf{A}_a(\mathbf{r}-\mathbf{r}_a(t)),
	\quad A_0=-\sum_{a=1}^N\mathbf{\dot r}_a\cdot\mathbf{A}_a(\mathbf{r}-\mathbf{r}_a(t))
	\ee
Note that $A$ \emph{is not} necessarily in the Coulomb gauge. It has the pleasant feature that a closed path in the space of controls 
$\{\mbf{r}_a\}$ is represented by a closed path of the potential, and hence a closed path of the Hamiltonian.

%%%%%%%%%%%%%%%%%%%%%%%%%%%%%%%%%%%%%%%%

%%%%%%%%%%%%%%%%%%%%%%%%%%%%%%%%%%%%%%%%
\subsection{The adiabatic evolution}\label{sec:ad}

We are interested in the evolution of the zero modes due to adiabatic motion of the fluxons. More specifically, 
we are  interested in the holonomy that describes the braiding of fluxons. This adiabatic problem has three subtle points: 
\begin{itemize}
\item  Gapless zero modes.  The zero modes lie at the threshold of the continuous spectrum so the adiabatic evolution is not 
protected by a gap. 
One may then appeal to  adiabatic theorems that cover the gapless case
\footnote{The intrinsic time scale can be determined by dimensional analysis and is: $mr^2/h$.  For point-like fluxons 
the only length scale is the distance between them.} \cite{avronElgart,Bornemann,AFGG}. These theorems 
hold provided the space of zero modes changes smoothly.   
Consider the collision of sub-critical point-like fluxons.  The norm of the fundamental solution $\braket{\psi_0}{\psi_0}$ 
diverges when they form a super-critical fluxon. It follows that the space of zero-modes does not behave smoothly upon flux collisions.  
Flux collisions then play the role of gap closure in  gapped adiabatic evolutions.  
Removing points of flux collisions endows the control space with a non-trivial topology.

  %%%%%%%%%%%%%%%%%%%%%%%%%%%%%%%%%
\item
Gauge freedom: Adiabatic phases are well defined (gauge invariant) for closed cycles of the Hamiltonian \cite{berry}. 
Braiding of fluxons is described by a closed cycle in control space $\{\mbf{r}_a\}$, and therefore also a closed cycle in the space of 
EM \emph{fields}, but not necessarily in  the space of Hamiltonian which depends on the potentials.   To correctly compute the holonomy, 
one needs the EM \emph{potentials} to make a closed cycle. (One \emph{is not} interested in phases that come simply 
from a change of gauge between the initial and final Hamiltonian.) This is taken care of by  the choice of gauge made in Eq.~(\ref{phi}). 
\item 
Parallel transport: 
In  standard adiabatic Hamiltonians \cite{teufel-Book} the {the adiabatic evolution is determined by the frozen Hamiltonian.  
This is not the case here where the weak electric field generates the evolution. 
It turns out to be  instructive   \cite{frohlich1993} to } 
write the Pauli evolution equation as
\be
iD_t\otimes \id\psi=\frac 1 {2m} \big((-i\nabla -e\mbf{A})\cdot\sigma\big)^2\psi,\quad D_t=\partial_t -ieA_0
\ee  
with $\partial_t$  replaced by the covariant time derivative $D_t$.
\end{itemize}

Let $\mbf{P}$ denote the spectral projection on the zero modes of  (the frozen) Pauli operator. The   evolution  generated by 
	\be\label{kato}
	i\dot\psi= \left(\mbf{ PH}_p\mbf{P}+i[\mbf{\dot P, P}] \right)\psi
	\ee  
maps unitarily $\Ran \mbf{P}\mapsto \Ran \mbf{P}$ \cite{Kato50}. The first term describes the action of the Hamiltonian on 
$\Ran \mbf{P}$ and the second term guarantees that the states remain within the instantaneous spectral subspace. 
Usually, the first term acts on $\Ran \mbf{P}$ as a c-number giving it just an overall phase and therefore in spite of being $O(1)$ 
it is usually less important than  the second which is only $O(\e)$.
For the case at hand, both terms act nontrivially on the zero modes space and are $O(\e)$.  Eq.~(\ref{kato}) reduces to
	\be
		i\dot\psi=\left(i[ \mbf{\dot P}, \mbf{P}] -e \mbf{P}A_0 \mbf{P}\right)\psi=\left(i[ \mbf{\dot P}, \mbf{P}] + 
e\sum_{a=1}^N \mathbf{v}_a \cdot \mbf{P} \mathbf{A}_a \mbf{P}\right)\psi
	\ee
And, we have used  Eqs.~(\ref{Pauli2},\ref{phi}). In particular, if $\psi$ and $\varphi$ are zero modes then $\psi=\mbf{P}\psi$ 
and $\varphi=\mbf{P}\varphi$ 
and the evolution of zero modes is governed by 
	\be\label{evolution1}
	i\braket{\varphi}{d\psi}=e \sum_a d\mathbf{r}_a \cdot \bra{\varphi} \mathbf{A}_a \ket{\psi}=e\sum_a\bra{\varphi}
    v_a  A_a+ \bar v_a \bar A_a\ket{\psi}dt
	\ee 
which is simply the parallel transport associated with the covariant derivative $D_t$. (Here $A_a\equiv (A_a)_z$.)

Given $\mathbf{A}$, the instantaneous fundamental solution $(\psi)_0$ is uniquely  determined  as in Sec.~\ref{sec:AC}. 
The adiabatic evolution can be viewed  as a rule for evolving the polynomial  $P(z,t)$
% which then fully determine the zero mode evolution
	\be\label{poly}
	\psi=P(z,t)\,\psi_0 =P(z,t)\,\prod_a(\psi_a)_0,
	 \quad P(z,t)=\sum_{j=0}^{D -1} p_j(t) z^j
	\ee
 %$\deg(p)< n-1$ .  When $deg(p)=0$ the $p=p(t)$ is a phase (and normalization) factor.  
In the rest of this section we show that the matrix elements of $\mbf{A}$ in the evolution equation (\ref{evolution1}) 
can be traded for the derivatives of the zero modes overlaps.  The main results are Eq.~(\ref{pt}), or equivalently Eq.~(\ref{gy876t}) below.

Note first that fundamental mode
 $(\psi_a)_0$ of each individual fluxon satisfies
	\be\label{kMode}
 	0=(\bar{\partial_z}-ie\bar A_a)(\psi_a)_0=-(\bar{\partial}_{a}+ie\bar A_a)(\psi_a)_0, \quad \partial_{a}=\partial_{\zeta_a}
	\ee
It follows from this that
	\begin{align} 
	e\bra{\varphi}  v_a  A_a+ \bar v_a \bar A_a\ket{\psi}=-i v_a\braket{\bar \partial_a\varphi} {\psi}+i\bar v_a \braket{\varphi}
{\bar\partial_a\psi}
	\end{align}
The evolution equation then takes the form
%%     \begin{align} \label{pt}
%%	0&=\braket{\varphi}{d\psi} +\sum   d\zeta_a  \braket{\bar\partial_a \varphi}{\psi}-
%%      d\bar \zeta_a  \braket{\varphi}{\bar\partial_a \psi}
%% \nonumber \\
%%	&=
%%	\bra{\varphi}d \log p\ket{\psi} +\sum d\zeta _a \partial_a\braket{\varphi}{\psi}
%%	\quad \psi=p(z,t)\psi_0,\quad \varphi=q(z)\psi_0
%%		\end{align}
%%
\be\label{pt0}	
0=\braket{\varphi}{d\psi} +\sum   d\zeta_a  \braket{\bar\partial_a \varphi}{\psi}-
d\bar \zeta_a  \braket{\varphi}{\bar\partial_a \psi} \ee
Using the fact that the fundamental modes $(\psi_a)_0$ evolve by rigid motion
%%%   \be
%%%   \der{\psi_0}{t}=\sum(\vec{v}_a\cdot\vec{\nabla}_a)\psi_0=\sum(v_a\partial_a+\bar{v}_a\bar{\partial}_a)\psi_0
%%%   \ee
%%%
\be
d{\psi_0}=\sum( d\mathbf{r}_a\cdot\mathbf{\nabla}_a)\psi_0=\sum(d\zeta_a\partial_a+d\bar{\zeta}_a\bar{\partial}_a)\psi_0
\ee
we finally arrive at the evolution equation for the polynomial {$P(z,t)$}:
\be\label{pt} 
  0=\bra{\varphi}d \log P\ket{\psi} +\sum d\zeta _a \partial_a\braket{\varphi}{\psi} ,
  \quad \psi={P}(z,t)\psi_0,\quad \varphi={Q}(z)\psi_0                               \ee
%%%%%%%%%%%%%%%%%%%%%%%%%
%%%%%%%%%%%%%%%%%%%%%%%%%
which may also be stated as:
\begin{claim}
The evolution of a zero mode under the change of the controls $d\zeta_a$ is determined by  the evolution 
of the corresponding polynomial $P(z,t)$. This is determined by the equations  
\be 
\langle\psi_0|\bar{Q}\,d{P}|\psi_0\rangle+\sum d\zeta_a\partial_a	\langle\psi_0|\bar{Q}\, P |\psi_0\rangle=0
\label{gy876t}
\ee
\end{claim}
%%%%%%%%%%%%%%

When the fluxons are pointlike subcritical $\Phi_a<1$  and the fundamental mode is chosen in the holomorphic gauge, as in 
Eq.~(\ref{holomorphic}), 
%This of course a special case. But, it turns out to be the interesting case, so let us look at it in some more detail.
the sum on the right of Eq.~(\ref{pt0})  
vanishes and the evolution equation simplifies to the statement that the velocity  in the manifold of zero modes vanishes:
\be\label{parallel}
0=\braket{\varphi}{d\psi}, \quad\quad \psi={P}(z,t)\psi_0,\quad \varphi={Q}(z)\psi_0
\ee
Note that the scalar product in the holomorphic gauge  is well defined even 
for fractional $\Phi_a$, independently of the way one chooses the cuts as long
as this choice is done consistently.

%%%%%%%%%%%%%%%%%
\subsection{The induced metric $\mbf{g}$} 
%%%%%%%%%%%%%%%%%%%
%%%%%%%%%%%%%%%%%%%
The $D$ dimensional space of zero modes can be naturally identified with the space of holomorphic polynomials  with $\deg(P)\leq D-1$.   
Natural coordinates are the coefficient $p_j(t)$ in $P(z,t)=\sum p_j(t) z^j$
\footnote{Any other basis in the space of  polynomials is legitimate.}.
Let 
\be
 p=(p_0,\dots, p_{D-1})^t ,\quad p\in\mathbb{C}^D
\ee
The  Hilbert space metric induces a metric on  $\mathbb{C}^D$
	\be\label{metric}
	(\mathbf{g})_{jk}= \braket{\psi_j}{\psi_k}, \quad \psi_j=z^j\psi_0, \quad j,k\in 0,\dots, D-1
	\ee
with the properties:
	\begin{itemize}
	\item $\mbf{g}$ is a positive,  hermitian,  $D\times D$ matrix.
	\item $\mbf{g}$  is gauge invariant:   It is independent of the choice of gauge for the (frozen) Pauli operator.  
	\item $\mbf{g}$ is a smooth  function of the fluxes, $\Phi_a$, provided $\Phi_T>D$. It blows up as the total flux 
$\Phi_T\searrow D$ reflecting the loss of one mode.
	\item When all the fluxons are finite, $R_a>0$, the metric is an everywhere smooth function of $\zeta$, the fluxon coordinates.
	\end{itemize}
{In the limit of pointlike} fluxons, $R_a\ll |\mathbf{r}_a-\mathbf{r}_b|$,
we can say more:
	\begin{itemize}
	\item The metric has an (approximate) block structure: All the (properly normalized) confined modes are in 
$1\times 1$ blocks, and all the free modes are in a single block. 
%\red{
(The terms connecting different blocks scale as positive powers of $R_a/|\mbf{r}_a-\mbf{r}_b|$.)
%}
	\item The block of the free zero modes is given by 
	\be\label{bkj}
	(\mathbf{g})_{jk}= \braket{\psi_{j}}{\psi_k}, \quad \psi_{j}(z;\zeta) =z^j \prod_a(z-\zeta_a)^{-\Phi_a'}, \quad 
\Phi_a'=\Phi_a-Max\{0,\lfloor\Phi_a\rfloor\} 
	\ee
	\item Under scaling the metric of the free modes behaves like: 
	\be\label{scaling}
	\mbf{g}_{jk}(\lambda \zeta)=\lambda^{k}\bar\lambda^j |\lambda|^{2(1-\Phi'_T)}\mbf{g}_{jk}(\zeta)
         \ee
         Note that the magnitude of $\lambda$ describes dilation and its phase a rigid rotation of the control space. 
\item  When two point-like fluxons $a$ and $b$ collide the metric blows up if $\Phi_a'+\Phi_b'\ge 1$.
\item It is natural to remove from control space (with coordinates $\zeta_a$) the points where $\mbf{g}=\infty$. This endows the 
control space of point fluxons with an interesting topology.

\end{itemize}
%%%%%%%%%%%%%%%%%%%%%%%%%%%%%%
\subsection{The adiabatic connection}\label{connection}
{Consider a path $\gamma: t\mapsto \zeta_a(t)$ in the space of controls. We are interested in the evolution of 
${p}(t)\in\mathbb{C}^D$ along the path.  } Making use of $\mbf{g}$, the transport equation, Eq.~(\ref{gy876t}),  takes the form
\be\label{asd}
0= \mbf{g} \,d {p} + \left(\sum d\zeta_a \partial_a \mbf{g}\right) \,{p} 
\ee
This can be written more compactly  using the  Dolbeault operator\footnote{$\partial$ stands for Dolbeault to be distinguished 
from  $\partial_z$ and $\partial_a$.} 
$\partial= \sum_{a=1}^N d\zeta_a \partial_a $ \cite{nakahara}
{(similarly $\bar{\partial}= \sum_{a=1}^N d\bar{\zeta}_a \bar{\partial}_a $)}:
	\be\label{yr}   
    	0= \mbf{g} \,d {p} +(\partial \mbf{g}) \,{p}     
	 \ee
The transport equation may then be written in terms of a connection$ {\cal A}$
 \be\label{gp}
0= (d+{\cal A})\,{p} , \quad {\cal A}= \mathbf{g}^{-1}( \partial \mathbf{g})
\ee

%%%%%%%%%%%%%%%%%%%%%
%%%%%%%%%%%%%%%%%%%%%
\subsection{The adiabatic curvature}
%%%%%%%%%%%%%%%%%%%%%%
The  connection determines the (adiabatic) curvature 2-form $\cal R$  by a standard   formula \cite{nakahara}:
\be\label{RR}
{\cal  R}=  d{\cal A}+{\cal A}\wedge{\cal  A}
=\bar{\partial}(\mbf{g}^{-1}\partial \mbf{g})
\ee
In the abelian case $D=1$ this simplifies into
\be\label{RA}  {\cal  R}=\partial\bar{\partial}\log\mbf{g} \ee
The curvature vanishes, ${\cal R}=0$, when the connection ${\cal A}$ is a pure gauge:
\be\label{puregauge}
{\cal A}=   \mathbf{ g}^{-1}_0d \mathbf{g}_0% , \quad d=\partial+\bar\partial 
\ee
The connection $\cal A$ of Eq.~(\ref{gp}) is ``half way'' to be a pure gauge ($\bar \partial$ is missing). It has special properties 
which we return to below.

%%%%%%%%%%%%%%%%%%%%%%%%%%%%%%%%%%%%%%%%
\section{The connection for point-like fluxons}

{The study of the connection for point like fluxons is easier in some cases and harder in others: 
It is relatively easy for the confined modes where  arguments based on the Aharonov-Bohm effect apply. 
It is more difficult in the more interesting case of  free  zero modes. 
This case too splits into  cases with different levels of difficulty: The abelian case is simpler than 
the non-abelian case, and rigid translations and rotations are easier than braiding.  This section is 
organized so that  we treat the easier and special cases first. A reader who prefers to start with the 
most general results may want to read the subsections below in different order.  }

\subsection{Block structure}\label{sec:block}
%%%   Mutual orthogonality of extended and localized states means that they belong to separate blocks  of $\mbf{g}$
%%%   and hence by Eqs(\ref{gp},\ref{RR}) also to  separate blocks of the connection and curvature forms.
%%%
By Eqs.~(\ref{gp},\ref{RR})
the connection $\cal A$ and curvature $\cal R$ inherit the block structure of $\mbf{g}$. 
As all the confined modes  are orthogonal to the free modes (in the limit $R\rightarrow0$) and to each other,
one concludes that the connection  $\cal A$ and curvature $\cal R$ split into a block  corresponding to all free states 
and a number of $1\times1$ blocks corresponding to each of the confined modes.  
Each block remains completely unaffected by other blocks and may therefore be discussed separately.

%%%%%%%%%%%%%%%%%%%%%%%%%%%%%%%%%%%%%%%%%%
\subsection{States confined to super critical fluxons   } \label{tight}

%%%%%%%%%%%%%%%%%%%%%%%%%%%%%%%%%%%%%%%%%%%%%%%%%%
Braiding the pair $(a,b)$ one expects\footnote{The result is approximate 
if $R_{a}$ is finite due to the power law tails of the state and exact in the limit $R_a\to 0$.  } a confined  mode on $a$ to  
acquire  Aharonov-Bohm phase $\nu_b=e^{-2\pi i \Phi_b}$.
Let us see how this can be understood from the machinery developed in Sec.~\ref{ch:adiabatic}.
Due to the  block structure of the connection, Sec.~\ref{sec:block},  the adiabatic evolution of a confined state has only 
an abelian phase factor $p(t)=e^{i\varphi(t)}$. 
From Sec.~\ref{super}  a mode confined at fluxon $a$ is described 
(e.g. in holomorphic gauge) by
$$\psi(z)=\psi_a(z-\zeta_a)\prod_{b\neq a}(z-\zeta_b)^{-\Phi_b'}\approx \psi_a(z-\zeta_a)\prod_{b\neq a}(\zeta_a-\zeta_b)^{-\Phi_b'}$$
where $\psi_a$ is independent of the $\zeta_b$s.
The metric ($1\times1$ block) is  the function
	\be
	\mbf{g} \approx \braket{\psi_a}{\psi_a} \, \prod_{b\neq a} |\zeta_a-\zeta_b|^{-2\Phi_b'}
	\ee
It follows from   Eq.~(\ref{yr})  that 
$$ id\varphi=d\log p=-\partial\log\mbf{g}=	d\left(\sum_{b\neq a}\Phi_b'\log(\zeta_b-\zeta_a)\right)$$
In particular, if $\zeta_b$ encircles $\zeta_a$ the phase is the Aharonov-Bohm phase $2\pi \Phi_b'$.

%%%%%%%%%%%%%%%%%%%%%%%%%%%%%%%%%%%%%
\subsection{Rigid translations and rotations of free states: $D\ge 1$}\label{ukhd}
%%%%%%%%%%%%%%%%%%%%%%%%%%%%%%%
%%  The results of  the previous subsection regarding the holonomy of  the free states under 
%%  translations and rotations can be generalized from the abelian case $D=1$ to the degenerate case, $D\ge 1$.    
 	\begin{itemize}
\item
Under (rigid) translations $d{\zeta}_a=\xi(t)\,dt $ of the entire fluxon configuration--independent of $a$
 but not necessarily of $t$--
the wave function undergoes simple translation $\psi(z)\rightarrow\psi(z-\xi dt)$.
{This fact may be verified by using Eqs.~(\ref{kernel1},\ref{evolution1}).}

\item
A rigid rotation of the entire configuration\footnote{Recall that individual fluxons are shifted parallel to themselves.}  
is described by (complex) scaling $d\zeta_a=\xi(t)\zeta_a dt$. From  Eqs.~(\ref{scaling}) we find
\[
\partial \mbf{g}_{jk}=(k+1-\Phi'_T) \mbf{g}_{jk} \xi\, dt
\]
From Eq.~(\ref{yr}) the $k$-th coefficient rotates independently of the others
\be
d{{p}}_k(t)=-(k+1-\Phi'_T) \, {p}_k(t) \,\xi(t) \,dt
\ee
In particular a full $2\pi$ rotation result in acquiring  of the (abelian) Berry phase 
\be\label{berry}
2\pi(\Phi'_T-k-1)
\ee
As $k$ is an integer this phase is identical (mod $2\pi$) for all free states.
(This could be anticipated from the fact that changing the origin of rotation mixes different $k$s.)

\end{itemize}
%%%%%%%%%%%%%%%%%%%%%%%%%%%%%%%%%%%%%%%%%%
%%%%%%%%%%%%%%%%%%%%%%%%%%%%%%

\subsection{A non-degenerate  free mode $D=1$}\label{non-deg}

\begin{figure}[htb]
    	\centering
    %\hskip -1 cm
   	\includegraphics[height=4 cm]{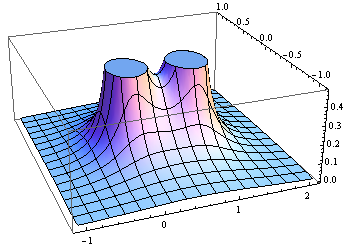}
	\caption{The adiabatic abelian curvature for 3 half fluxes. It is singular when fluxon collides, namely 
at $u=1$ and $u=0$.  It is symmetric under $u\mapsto \bar u$. See Eq.~(\ref{fig:r}) and Appendix ~\ref{3flux} for 
more details on how the figure was drawn.}
\label{fig:curvature}
	\end{figure}

%The study of the holonomy of the free modes is easier in the abelian case where  $D_f=1$. 
Consider $N$ subcritical fluxons 
$\Phi_a<1$. The metric $\mbf g$ is simply the  norm of the fundamental mode
	 \be
    	 \mbf{g}=\braket{\psi_0}{\psi_0}, \quad \psi_0(z;\zeta)=\prod_a(z-\zeta_a)^{-\Phi_a}
   	 \ee
   	 \begin{exa}
Suppose $N=3$, holding $\zeta_1=0, \zeta_2=1$ fixed. The  adiabatic curvature in the remaining  
$u=\zeta_3$ coordinate is given  by Eq.~(\ref{RA}) 
	\be\label{curva}
	{\cal R}=d{\cal A}=\bar \partial_u\partial_u\log \mbf{g}(u),\quad
\mbf{g}(u)= \int \frac{d^2z}{|z|^{2\Phi_1} |z-1|^{2\Phi_2}|z-u|^{2\Phi_3}}%\nonumber\\ 
	\ee
There is no reason for $\log\mbf{g}(u)$ to be a harmonic function. In Appendix \ref{3flux} we show that 
$\mbf{g}$ and ${\cal R}$ can be evaluated in terms of hyper-geometric functions.   Indeed  as one can see from 
Fig. \ref{fig:curvature} for three half fluxes the curvature does not vanish.
 \end{exa}
%	\item
When $N=2$  and $D=1$  braiding is topological by a special reason that we shall discuss in Sec. \ref{sec:square}. 
However,   for $N\ge 3$ and $D=1$, the braiding of two fluxons $(a,b)$, keeping all the other fluxons fixed   is, in general, 
path  dependent.

%%%%%%%%%%%%%%%%%%%

 \begin{figure}[htb]
    	\centering
    %\hskip -1 cm
   	\includegraphics[height=6 cm]{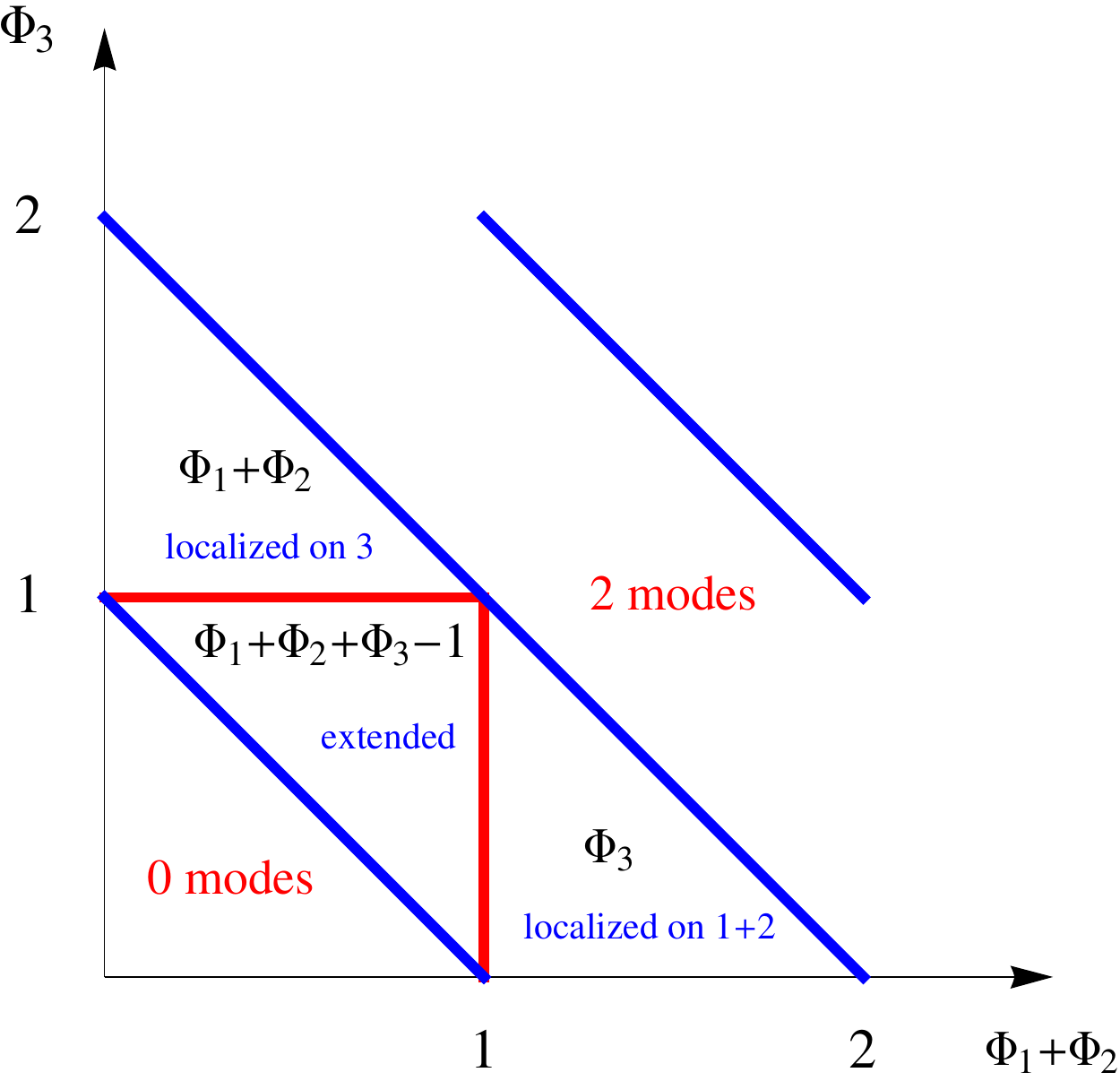}
	\caption{A ``phase diagram'' for the holonomy of fluxon 3 going around the pair of nearby fluxons $1+2$
{i.e. $r_{12}\ll r_{13}$}.  
The diagonal blue lines delineate  regions without zero modes, with one zero mode and with 2 zero modes.
In the triangle marked by $\Phi_3$ the charge is localized on $1+2$ fluxons and  Berry's phase is $2\pi \Phi_3$ as one would 
expect from the Aharonov-Bohm effect. The triangle marked  $\Phi_1+\Phi_2$ describes a zero mode confined on 3 and the phase 
is  $2\pi(\Phi_1+\Phi_2)$. In the triangle marked by $\Phi_1+\Phi_2+\Phi_3-1$ the state is free and the Berry phase is
 $2\pi(\Phi_1+\Phi_2+\Phi_3-1)$.}  
	\end{figure}
%%%%%%%%%%%%%%%%%%%%

%\item In particular, for two fluxons the holonomy is topological, and is a special case of the formula above.

%%%%%%%%%%%%%%%%%%%%%%%%%%%%%%%%%%%%%

\subsection{The  connection for  free states of  point-like fluxons }\label{sec:connection1}

For the sake of simplicity of the notation we shall  write $\Phi_a=\Phi_a'<1$
 and  $D=D_f$ in this section.

In  Appendix \ref{sec:integrals}  we show that the metric $\mbf{g}$ for the free zero modes  factorizes into a holomorphic 
and anti-holomorphic factors. By Eq.~(\ref{gexplicit}):
\be\label{kjdc}
\mathbf{g}(\zeta;\Phi)= \mathbf{\Psi}^* (\zeta;\Phi)\mathbf{G}(\Phi) \mathbf {\Psi}(\zeta;\Phi)
\ee
where:
\begin{itemize}
\item  $\mathbf{\Psi}(\zeta,\Phi)$ is    an $N\times D$ holomorphic (matrix) function whose matrix elements are given 
{ in Eq.~(\ref{paj}),  as}
\be\label{paj1}
\mathbf{\Psi}_{ak} (\zeta)= \int_{\xi_0}^{\zeta_a} d\xi\, \psi_k(\xi), \quad a\in 1,\dots,N, \quad k\in 0,\dots, D-1 
\ee
%\be\label{paj}
%\mathbf{\Psi}_{aj}(\zeta)= \Psi_j(\zeta_a;\zeta ,\xi_0) %\quad \Psi_j(z;\zeta,\xi_0)=\int_{\xi_0}^z \frac{x^j dx}{(x-\zeta_1)^{\Phi_1}
%%\dots(x-\zeta_N)^{\Phi_N}}
%\ee

%\item $\mbf{\Psi}_{aj}$ is finite 
$\mathbf{\Psi}(\zeta,\Phi)$  involves a choice, which we call a {\it  gauge choice}, and is hidden in the freedom of $\xi_0$.  
Alternatively one could add arbitrary integration constant $c_k$ to the r.h.s.

\item  $\mathbf{G}(\Phi)$ is an $N\times N$ hermitian matrix which is independent of the controls 
$ \zeta\in\mathbb{C}^N$.  
Its explicit form is given in appendix \ref{sec:integrals}.
It has rank $N-1$ with its kernel generated by the vector $1_N=(1,1,...1)^t$.
Note that this guarantees that the metric $\mathbf{g}$ is not affected by adding  arbitrary integration constant to Eq.(\ref{paj1}).
%%    \be
%%    \mbf{G}_{ab}=\frac{\sin\left(\pi\Phi_a \right)}{\sin\left( \pi\Phi_T\right)}\times
%%    \begin{cases}
%%    -\sin\left(\pi(\Phi_T-\Phi_a)\right)& a=b\\
%%     \sin\left(\pi\Phi_b\right)
%%     \exp\left[i\pi\left(\Phi_T-\sum_{c=a}^{b-1}(\Phi_c+\Phi_{c+1})\right)\right]&a<b
%%    \end{cases}          \ee

\end{itemize}
%%%%%%%%%%%%%%%%%%%%%%%

%%%%%%%%%%%%%%%%%%%

Substituting the expression (\ref{kjdc}) for the metric into Eq.~(\ref{yr}) leads to the transport equation 
\be    
\underbrace{{\mathbf{ \Psi}}^*\, {\mathbf{G}}}_{D\times N}\,  
{d}\left({\mathbf{\Psi}}{p}\right)=0     \label{hjs} \ee
These are $D$ equations for the evolution of the $D$ coefficients $p\in\mathbb{C}^{D}$. The solutions of these equations are, 
in general, path dependent.
{If one could cancel out the left factor $\mathbf{ \Psi}^* \mathbf{G}$, it would follow that the holonomy is topological.
In general however this cannot be done since $\mathbf{ \Psi}^* \mathbf{G}$ is not an invertible (or even a square) matrix.}
 The holonomy is, in general, not topological.

%%%%%%%%%%%%%%%%%%%%%%%%%%%%%%
\subsection{Topological holonomy}\label{sec:square}
When the number of free  zero modes is maximal
	\[
	D_f=N-1
	\]  
 the holonomy is topological. 
\begin{exa}

The simplest example of this kind is  $N=2$ and $D_f=1$.  The Berry phase associated with taking one fluxon around the other is 
{\it topological} and is given by Eq.~(\ref{berry}) with $k=0$
\[
2\pi(\Phi_T-1)
\]
\end{exa}
%	\item 

One way to show that this is the case is by using the  `gauge' freedom  to choose $\xi_0=\zeta_N$
in Eq.(\ref{paj1}) so that $\mathbf{\Psi}_{Nj}=0$ for all $j$. 
By throwing its last row we can then view $\mathbf{\Psi}$ as a square $D\times D$ matrix which we 
denote $\mbf{\Psi}_\square$.
%%%%%%%%%%%%%%%%%%%%%%%%%%%%%%%%%%
We may then write for the metric
\be\label{square1}
\mbf{g}_{jk} = \sum_{a,b=1}^{D} \bar{\mathbf{\Psi}}_{aj} G_{ab} \mathbf{\Psi}_{bk}
 \Longleftrightarrow \mathbf{g}= \mbf{\Psi}^*_\square \mbf{G}_\square  \mbf{\Psi}_\square
 \ee 
where $\mbf{\Psi}_\square, \mbf{G}_\square, \mbf{\Psi}^*_\square$ are all square $D\times $D matrices. 
{Repeating the arguments of the previous subsection Eq.~(\ref {hjs}) now takes the form}
\be    \mbf{\Psi}^*_\square \mbf{G}_\square {d}\left({\mathbf{\Psi}}_\square  p \right)=0     \ee
Since $\mathbf{g}$ is positive, all  of its ($D\times D$) factors are invertible.
The equations of parallel transport therefore reduces into $D$ conservation laws and ${\cal A}$ is a pure gauge :
\be \label{ksg}
d( \mathbf{\Psi}_\square\,  p)=0, 
\quad {\cal A}= \mathbf{g}^{-1} \partial \mathbf{g}=  
%%  \mbf{\Psi}^{-1}_\square\partial \mbf{\Psi}_\square=  
\mbf{\Psi}^{-1}_\square d \mbf{\Psi}_\square
\ee    
It follows that the holonomy of adiabatic transport is determined by the monodromy of 
the (multivalued function) $\mbf{\Psi}_\square$.

{The gauge choice $\xi_0=\zeta_N$ has the disadvantage of treating $\zeta_N$ on different footing than 
the other $\zeta_a$s. Moreover, in braiding operations in which the $N$th fluxon is more than a spectator,
any dependence of $\xi_0$ on $\zeta_N$ can lead to extra complication.
For this reason we shall prefer in the next section to fix $\xi_0$ to be a constant independent of $\zeta$.} 

\begin{rem}
{One may avoid the `gauge' choice $\xi_0=\zeta_N$ and rewrite}
equation (\ref{ksg})  in a gauge independent way as
	\be\label{dgij}      
	    d( \mathbf{\Psi}\,{p})\in\mathbb{C}\,\mbf{1}_N      
         \ee 
where $\mbf{1}_N=(1,1,...1)^t$ is the generator of $\ker(\mbf{G})$
{and the complex coefficient on the r.h.s depends on the arbitrary integration constant chosen in the definition of
$\mathbf{\Psi}$ }.
\end{rem}

%%%%%%%%%%%%%%%%%%%%%%

\section{Fluxons braiding, non-abelian unitaries  and  anyons}\label{sec:nonabelain}
 We have seen that when $D_f=N-1$  there is no curvature. 
Hence, if the unitary holonomy operators are non trivial, and non-abelian, 
then fluxon braiding can be viewed as (non abelian) topological unitary operations on the manifold  of zero modes. 
We start by computing the monodromy of braiding of distinct fluxons.  In the special case that the fluxons carry identical fluxes, 
they may be viewed  {as identical} anyons.  In particular, when the fluxes are identical, it obeys the braiding rules of Burau 
representation of the braid group.

%%%%%%%%%%%%%%%%%%%%%%%%%%%%
%%%%%%%%%%%%%%%
\begin{figure}[h!]
\hspace{3 cm}
	\begin{picture}(300,160)(0,0)
	\thicklines
	\color{red}
	\put(30,40){\line(1,0){200}}
	\put(30,40){\circle*{5}}
	\put(80,90){\line(1,0){150}}
	\put(80,90){\circle*{5}}
	\put(47,140){\line(1,0){183}}
	\put(47,140){\circle*{5}}
	\color{black}
	\put(25,25){$\zeta_1$}
	\put(125,45){$\Sigma_1$}
        \put(220,65){$\infty_1$}
	\put(75,75){$\zeta_2$}
	\put(135,95){$\Sigma_2$}
	\put(220,105){$\infty_2$}
	\put(42,125){$\zeta_3$}
	\put(92,145){$\Sigma_3$}
	\put(220,155){$\infty_3$}
	\put(220,20){$\infty_0=\infty_3$}
	\put(7,10){\vector(1,0){240}}
	\put(7,10){\vector(0,1){150}}
	\put(230,-4){$x$}
	\put(-9,149){$y$}
	\end{picture}
\caption{
The convention for ordering the cuts is shown in the case of three point-like fluxons located at $\zeta_1,\zeta_2,\zeta_3$. 
With each fluxon one associates a cut $\Sigma$. The cuts define
three different regions that extend to infinity. {The cuts are ordered so that as one goes counter-clockwise along a big circle
(near infinity) the cuts are traversed successively according to their $a$-indexing.  We identify $\Sigma_{N+1}$ with $\Sigma_1$}.
The function  $\mathbf{\Psi}$ defined in Eq.~(\ref{ppppsi}), takes different limiting values at $\infty_a$.}\label{fig:cuts}
%\end{center}
\end{figure}
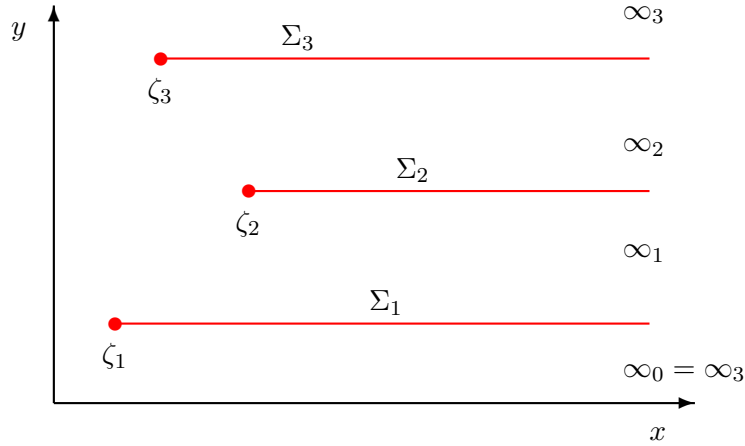

%%%%%%%%%%%%%%%

%%%%%%%%%%%%%%%%%

%%%%%%%%%%%%%%%%%%%%%%%%%%%%%%%
\subsection{The monodromy of braiding fluxes}

We start by computing the monodromy of $\mbf{\Psi}$. 
In subsection \ref{uyufc} we shall relate it to the holonomy of the adiabatic evolution.

%We are interested in the monodromy of $\mathbf{\Psi}$ as fluxon j goes around fluxon k.  
The components of the matrix $\mathbf{\Psi}$ are by { Eq.~(\ref{paj1}),}
	\be\label{ppppsi}
	\mbf{\Psi}_{aj}(\zeta)=\int_{\xi_0}^{\zeta_a} d\xi\  \xi^j \prod_{b=1}^N (\xi-\zeta_b)^{-\Phi_b}
	\ee
Since $j$ is an integer the factor $\xi^j$ in the integrand has no interesting effect on 
the monodromy and we can {ignore the index $j$}
(and henceforth drop it) without risk.  
%%%%%%%%%%%%%%%%

%%%%%%%%%%%%%%%%%%%%%%%%%%%%%%%%%

\begin{figure}[htb]
    	\centering
    %\hskip -1 cm
%%%%%%%%%%%%%%%%%CCCCHHHH%%%%%%%%%
        \includegraphics[width=13 cm]{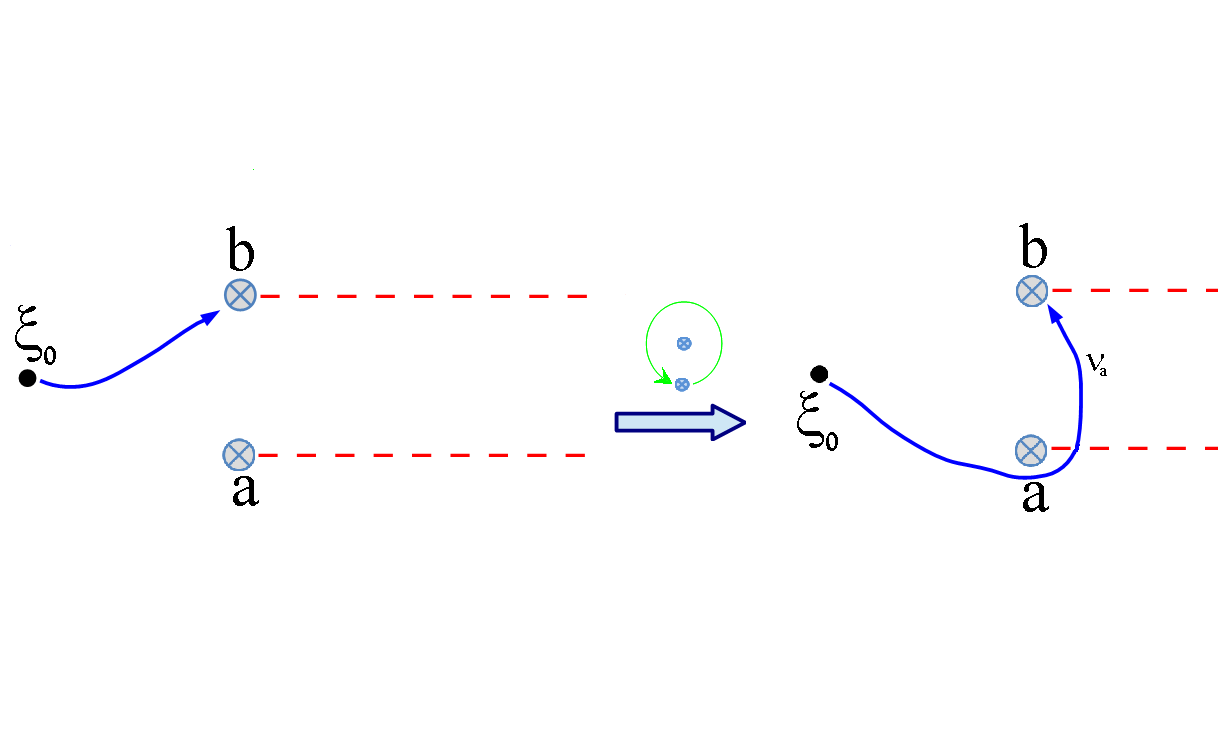}
%%%%%%%%%%%%%%%%%CCCCHHHH%%%%%%%%%
	\caption{The figure shows fluxon $a$ with its (red dashed) branch cut $\Sigma_a$ going counter clock-wise  
around fluxon $b=a+1$ with its (red dashed) branch cut $\Sigma_b$.  As $a$ encircles $b$ 
	the integration path (blue) from $\xi_0$ to $\zeta_b$ loops around the branch 
point of the cut $\Sigma_a$.  
The integration from $a$ to $b$ in the old and new paths are related by a factor $\nu_a$.}\label{fig:braidB1} 
	\end{figure}

\begin{figure}[htb]
    	\centering
    %\hskip -1 cm
%%%%%%%%%%%%%%%%%CCCCHHHH%%%%%%%%%
   	\includegraphics[width=13 cm]{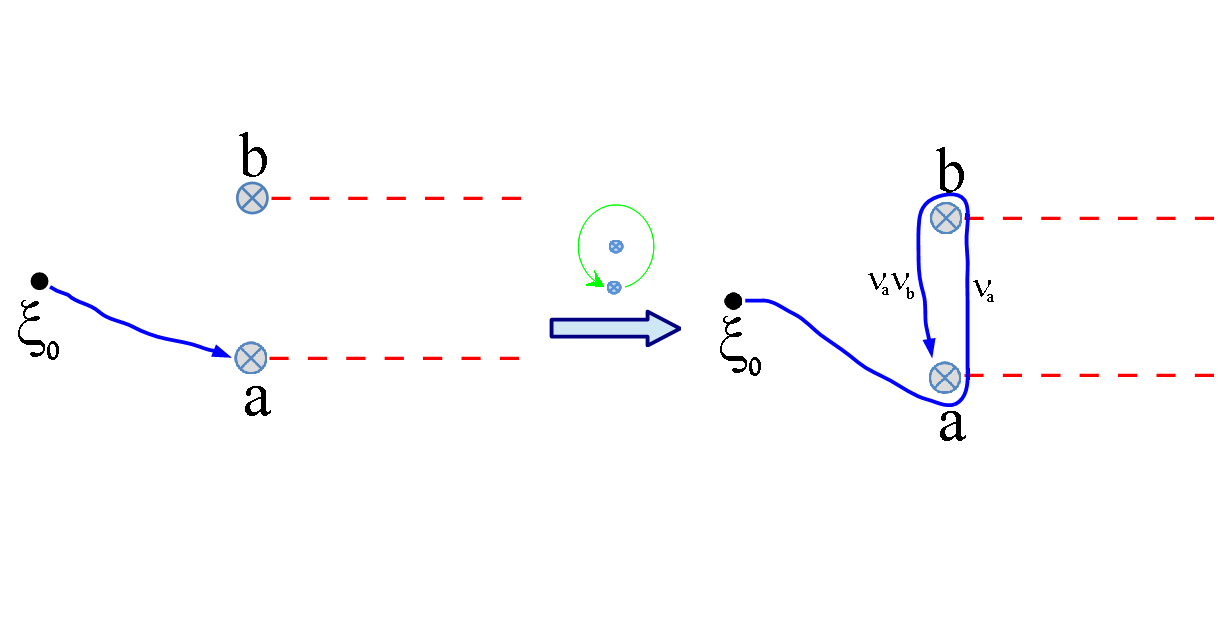}
%%%%%%%%%%%%%%%%%CCCCHHHH%%%%%%%%%
	\caption{The deformation of the path of integration from the fiducial point $\xi_0$ to the fluxon at $\zeta_a$.
	As the $a$-th fluxons encircle b counter-clockwise it ``stitches'' $b$. The old and new values of 
$\Psi_a$ differ by integrations from $a$ to $b$ along two sides of the two cuts.
	}\label{fig:braidA2}
	\end{figure}

Choosing integration paths from $\xi_0$ to $\zeta_1,\dots,\zeta_N$ which do not cross any of the cuts shown in Fig. \ref{fig:cuts}
leads to a standard definition of $ \Psi=(\Psi_1,\dots,\Psi_N) ^t$.
Upon cyclically moving the fluxon positions
\footnote{Here, unlike in section \ref{sec:square}, we keep fixed $\xi_0$ independent of $\zeta$.
If we do otherwise the monodromy would get extra contribution from possible movement of $\xi_0$.}
$\zeta_a$ these paths are deformed as seen  in Figs. \ref{fig:braidB1},\ref{fig:braidA2} into
paths which typically do cross the cuts. This leads to another branch $ \Psi'=(\Psi_1',\dots,\Psi_N') ^t$ of the multivalued function.
The monodromy  is  an $N\times N$ matrix  $\mbf{M}$:

%%%%%%%%%%%%%%%%%%%%%%%%%%%%%%%%%
	 \be\label{M}
 	\Psi'=\mbf{M}\Psi, \quad \Psi=(\Psi_1,\dots,\Psi_N)^t
 	\ee
It is useful  to collect properties of $\mbf{M}$ before one actually computes it, as they provide tests on the computations. 

\begin{itemize}
\item 
Since the fluxons may be moved backwards,  the monodromies $\mbf{M}$ must be invertible and 
generate a group. 
\item  Since $\mathbf{g}$  {(of Eq.~(\ref{kjdc}))} must not be  affected by the monodromy,  
$\mbf{M}$ and $\mathbf{G}$ must satisfy a consistency condition
	\be\label{unitary}
	\mathbf{G} = \mbf{M}^* \mathbf{G} \mbf{M}
	\ee 
	(This may be viewed as a unitarity condition). 
\item {Adding an integration constant (or equivalently changing $\xi_0$) in Eq. (\ref{ppppsi}) corresponds to
$\Psi\mapsto\Psi+c\mbf{1}_N$ and $\Psi'\mapsto\Psi'+c\mbf{1}_N$.  Consistency with Eq.(\ref{M}) thus requires}
\[
\mbf{M}\,\mbf{1}_N=\mbf{1}_N
\]
\end{itemize} 
%%%%%%%%%%%%%%%%%%
%%%%%%%%%%%%%%%%

{A proper definition of the monodromy of $\Psi$ requires choosing some definite convention for placements of the cuts, 
see Fig.~\ref{fig:cuts}.
We shall take the cut $\Sigma_a$ to run from $\zeta_a$ to $\infty$, and we order them in such a way that as one goes 
counter-clockwise along a big circle (near infinity) the cuts $\Sigma_a$ are traversed successively according to their 
$a$-indexing.}

Let us now compute the monodromy as the flux ${a}$ encircles an adjacent flux 
${b}=a+1$ counter-clockwise.  The integration path  {associated with $\Psi_b$ loops around the branch point $a$} as shown in 
Fig. \ref{fig:braidB1}.
The {deformed path gives}
\begin{align}
(\Psi')_b&=\int_{\xi_0}^{\zeta_a}\psi(z)dz+\nu_a\int_{\zeta_a}^{\zeta_b}\psi(z)dz\nonumber \\
&=\Psi_a+\nu_a(\Psi_b-\Psi_a)\label{monodb}
\end{align}

By similar considerations as $a$ loops around $b$,  the integration path {associated with}
$\Psi_a$ ``stitches'' $b$ as shown in Fig. \ref{fig:braidA2}.
It follows that 
\begin{align}\label{84}
(\Psi')_a&=\int_{\xi_0}^{\zeta_a}\psi(z)dz+\nu_a\int_{\zeta_a}^{\zeta_b}\psi(z)dz
+\nu_b\nu_a\int_{\zeta_b}^{\zeta_a}\psi(z)dz\nonumber\\
&=\Psi_a+ \nu_a(\Psi_b-\Psi_a) + \nu_a\nu_b(\Psi_a-\Psi_b)\nonumber\\
&=\big(1-\nu_a+\nu_a\nu_b\big)\Psi_a+\nu_a(1-\nu_b)\Psi_b
\end{align}
%%%%%%%%%
%%%%%%%%%%%%%%
All other components of $\Psi$ remain unaffected.
The $2\times 2$ nontrivial block of the monodromy matrix is therefore
\be
\mbf{M}(\nu_a,\nu_b)={\left(\begin{array}{cc}
1-\nu_a+\nu_a\nu_b & \nu_a(1-\nu_b) 
\\
1-\nu_a  & \nu_a
  \end{array}\right )}, \quad \det \mbf{M}=\nu_a\nu_b
 \ee
The monodromy matrix  is not symmetric in $a,b$  due to our convention for ordering the cuts {counter clockwise}.
%\margin{I do not understand}
   $\mbf{M}(\nu_a,\nu_b)$ is related to  $\mbf{M}(\nu_b,\nu_a)$ by
\[
\mbf{M}(\nu_b,\nu_a)= \sigma_x \mbf{M}^{-1}(\bar\nu_a,\bar\nu_b) \sigma_x
\]
By Eq.(\ref{unitary}) the spectrum of $M$ should lie on the unit circle. 
\footnote{ $ker(\mathbf{G})$ is spanned by $\mbf{1}_N$ known to be an eigenvector of $M$.}

Indeed:
 \be\label{evM}
 Eigenvalues(\mbf{M})=\{1,\nu_a \nu_b\}
 \ee
%%%%%%%%%%%%%%%%%%%%%%%%%%%%%%%%%%
%%     \begin{exa}
%%     In the case of three identical fluxes
%%     \be
%%     \mbf{M}_{12}(\nu)={\left(\begin{array}{ccc}
%%     1-\nu +\nu^2 & \nu(1-\nu)&0 \\
%%     1-\nu & \nu&0 \\  0&0&1
%%      \end{array}\right )}
%%      \ee
%%      $\mbf{M}_{12}$ satisfies the unitarity condition Eq.~(\ref{unitary}), with $\mathbf{G}$ 
%%      given in Eq.~(\ref{GN}).   $\mbf{M}_{12}(\nu)$ has eigenvalues $(1,1,\nu^2)$ while $\mbf{M}_{12}(\nu)\mbf{M}_{23}(\nu)$  
%%      has eigenvalues $(1,\nu,\nu^3)$. 
%%      \end{exa}\margin{so ?}
%%%%%%%%%%%%%%%%%%%%%%%%%%%%%%%%%%%%%%%%%%
\subsection{Braiding identical fluxes}\label{sec:burau}
 \begin{figure}[htb]
    	\centering
    %\hskip -1 cm
%%%%%%%%%%%%%%%%%CCCCHHHH%%%%%%%%%
   	\includegraphics[width=13 cm]{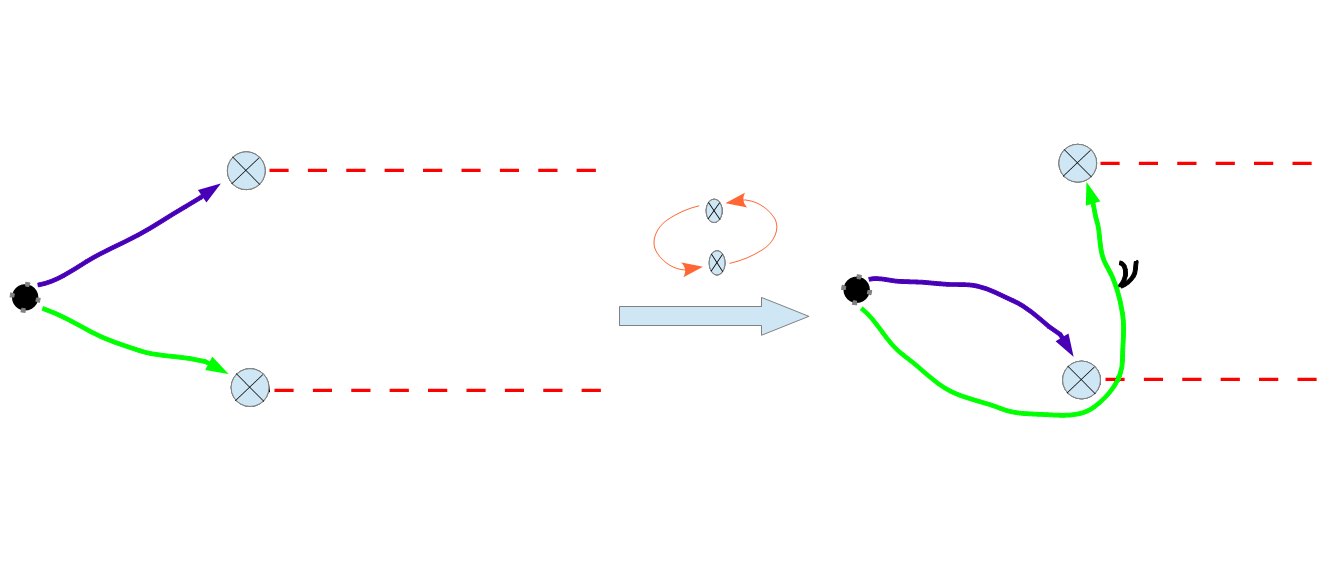}
%%%%%%%%%%%%%%%%%CCCCHHHH%%%%%%%%%
	\caption{Exchanging identical fluxes}\label{fig:ex}
	\end{figure}

In the special case where all fluxons are identical (having the same
$\Phi_a$) it makes sense to consider
also a permutation of two adjacent fluxons. This leads to the Burau representation
of the braid group \cite{birman}.

Indeed, inspecting Fig.~\ref{fig:ex} we see
\begin{align}
(\Psi')_a &= \ \Psi_a+\nu(\Psi_b-\Psi_a)\\
(\Psi')_b&= \Psi_a \nonumber
\end{align}
The monodromy matrix 
has the single nontrivial $2\times2$ block
\be\label{ex}
\mbf{M}(\nu)={\left(\begin{array}{cc} 
1-\nu & \nu\\ 
 1 & 0
\end{array}\right )}, \quad \det\, \mbf{M}=-\nu
\ee
with eigenvalues $\{1,-\nu\}$. 
{Note that when $\nu\rightarrow1$ this reduces to the standard representation of the symmetric group.
This may be understood as due to the fact that in this limit the free zero-modes turn into confined modes
which move with the fluxons. 
The case $\nu=-1$ (i.e. $\Phi_a={1\over2}$)   does not occur since it is inconsistent
with the assumption $D_f=N-1$,  {see  Eq.~(\ref{con}). } 
%%%%%%%%%%%%%%%%
%%%%%%%%%%%%%%%%

\subsection{The non-abelian holonomy}\label{uyufc}
%%%%%%%%%%%%%%%%%%%

The (non-abelian) holonomy $\mbf{U}(\gamma,\zeta)$ for a closed path $\gamma$ and base point $\zeta$,
acts unitarily on the $D_f=N-1$ dimensional space of (free) zero modes at $\zeta$:
\[
\mbf{U}:\sum p_j|\psi_j\rangle\mapsto\sum p_j'|\psi_j .\rangle
\]
One may write $\mbf{U}|\psi_j\rangle=\sum u_{ij}|\psi_i\rangle$ or equivalently $p_i'=\sum u_{ij}p_j$. 
Since the basis $\{|\psi_j\rangle\}$ is not orthonormal, the matrix $\mbf{u}$ is not unitary.
Instead it satisfies $\mbf{u}^* \mbf{g} \mbf{u}=\mbf{g}$. 
This is consistent with unitarity of the holonomy operator
$$\mbf{U}=\sum |\psi_j\rangle\left( u g^{-1}\right)_{ji}\langle\psi_i|$$

The previous sections make it clear that on the $D_f=N-1$ dimensional space of free zero modes,
the $(N-1)\times(N-1)$ matrix $u$ is closely related to the $N\times N$ monodromy matrix $M$. 
The exact relation is however complicated by the 'gauge' freedom
of fixing $\xi_0$. We would like in this section to state this relation more precisely.

For simplicity, consider first only braidings which do not involve the $N$-th fluxon.
Using the 'gauge' choice $\xi_0=\zeta_N$ one writes the conservation law, Eq.~(\ref{ksg}), taking ${p}$ around a 
closed path (based at $\zeta$):  
\be
\mbf{\Psi}_\square\,{p}=\mbf{ \Psi'}_\square \,{p'}= \mbf{M}_\square\mbf{\Psi}_\square \,{p'}
 \ee
 ($\mbf{M}_\square$ is the $D\times D$ matrix obtained from $\mbf{M}$ by deleting its last row and column.) 
The last identity used the definition of $\mbf{M}, $ Eq. (\ref{M}).
Hence
  \be\label{u}
  p_i'=\sum_j u_{ij}p_j,\quad 
\mbf{u}=(\mbf{ \Psi}_\square)^{-1}\mbf{M}^{-1}_\square\mbf{\Psi}_\square
 \ee

The derivation given above was special to the case where $\zeta_N$ was a spectator. 
Below we give an analysis of the general case.

The monodromy matrices $M$ acting on $\mathbb{C}^N$ preserve the vector $\mbf{1}_N\in\mathbb{C}^N$ 
i.e. satisfy $M\mbf{1}_N=\mbf{1}_N$.
It follows that $M$ defines a linear transformation $\tilde{M}$ on the quotient space $V_0=\mathbb{C}^N/\mathbb{C} \,\mbf{1}_N$.
We shall show that the holonomy $\mbf{U}(\gamma,\zeta)$ for a closed path $\gamma$ and base point $\zeta$
is obtained from $\tilde{M}$ by a similarity transformation.

The hermitian matrix $G$ satisfies $G\mbf{1}_N=0$. Therefore the  hermitian form it defines on $\mathbb{C}^N$, 
projects to a hermitian form $\tilde{G}$ on $V_0=\mathbb{C}^N/\mathbb{C} \,\mbf{1}_N$. Moreover since $G$ has $D$
positive eigenvalues the form $\tilde{G}$ must give a (non degenerate) inner product on $V_0$.
Eq.~(\ref{unitary}) shows that $\tilde{M}$ are unitary relative to this inner product on $V_0$.

By Eq.(\ref{dgij}) and the definition of the monodromy $M$ one has
$$M\mbf{\Psi}p'=\mbf{\Psi}p\;\mod\, \mbf{1}_N$$
Recall that $\mbf{\Psi}$ is an $N\times(N-1)$ matrix i.e. a map $\mathbb{C}^{N-1}\rightarrow\mathbb{C}^N$. 
Denoting by $\tilde{\mbf{\Psi}}$ the corresponding map into $V_0=\mathbb{C}^N/\mathbb{C} \,\mbf{1}_N$
we conclude 
$$\tilde{M}\tilde{\mbf{\Psi}}p'=\tilde{\mbf{\Psi}}p$$
As $\tilde{\mbf{\Psi}}$ is clearly invertible (as follows e.g. from $\tilde{\mbf{\Psi}}^*\tilde{G}\tilde{\mbf{\Psi}}=g$),
we see that
$$u=\tilde{\mbf{\Psi}}^{-1}\tilde{M}^{-1}\tilde{\mbf{\Psi}}$$
In particular the eigenvalues of the holonomy  $\mbf{U}$ of fluxon braiding are related to the eigenvalues of the monodromy $M$ by 
\be\label{MU}
Eigenvalues(\mbf{M})=Eigenvalues(\tilde{\mbf{M}})\cup\{1\}=Eigenvalues(\mbf{U}^{-1})\cup\{1\}
\ee
It follows from the results of the previous sections  that when the fluxon $a$ 
goes around the fluxon $b$, the holonomy matrix eigenvalues are
\be\label{main2}
Eigenvalues(\mbf{U})= \{1, \bar{\nu}_a\bar{\nu}_b\}
\ee

\begin{rem}
By considering the $N\times N$ matrix $\Lambda(\zeta)={\left(\begin{array}{cc} \mathbf{\Psi} & \mbf{1}_N \end{array}\right )}$
obtained by adding $\mbf{1}_N$ as an extra column to $\mathbf{\Psi}(\zeta)$ and defining
\[
|a\rangle=\sum_{j=0}^{D-1}{\Lambda}^{-1}_{ja}|\psi_j\rangle,\, a=1,...N
\]
we can obtain  the simple relation $\mbf{U}|a\rangle=(\mbf{M}^{-1})_{ba}|b\rangle$
{as well as $\langle a|b\rangle=G_{ab}$}. The set $\{|a\rangle\}$ is however an over-spanning set
rather than a basis as it satisfies $\sum |a\rangle=0$.  %%\margin{}

\end{rem}

%%%%%%%%%%%%%%%%%%%%%%

%%%%%%%%%%%%%%%%%%%%%%%%%%%%%%%%%%%%%%

%%%%%%%%%%%%%%%%%%%%%%%%%%%

%%%%%%%%%%%%%%%%%%%%%%%%%%%%%%%%%%%%

\thanks{{\bf Acknowledgement:} The research is supported by ISF. We thank M. Fraas, Y. Aharonov, N. Lindner, A. Ori, N. Cohen 
and P. Seba for discussions.}

\appendix

\section{Factorization of the metric}\label{sec:integrals}
%%%%%%%%%%%%%%%%%%%%%%%%%%%%%%%%%%

In this section we show that the metric $\mbf{g}$ for the free modes of point-like fluxons factorizes into a product of 
a holomorphic and anti-holomorphic factors. Since we are interested only in the free states,  we will,
for  notational simplicity, assume all fluxons are subcritical. If this is not the case one should
replace $\Phi_a$ by its fractional part. 

Let $\mbf{\Psi}_j$ denote the primitive integral of $\psi_j$:
\be\label{F}
 \Psi_j(z;\zeta,\xi_0)=\int_{\xi_0}^z d\xi\, \psi_j(\xi), \quad\psi_j(\xi)=\xi^j\prod_a (\xi-\zeta_a)^{-\Phi_a},\quad j=0,\dots,D-1
\ee
We shall refer to the choice of $\xi_0$  as a choice of a gauge. 

For $N=2,3$ Mathematica can evaluate $\Psi_j$ of Eq.~(\ref{F}) in terms of known special functions. In general, when $N\ge 4$ the 
integral form is the best we can do.

Since $ dx\wedge dy=\frac i 2 dz\wedge d\bar{z}$  one of the two integrations in Eq.~(\ref{bkj}) for the metric is  for free 
\be\label{g}
\mbf{g}_{jk}=\frac i 2 \int d {\Psi}_k\wedge  d\bar {\Psi}_{j}=\frac i 2 \sum_{a}  \int_{\Sigma_a} {\Psi}_k  {\bar\psi}_{j} d\bar z
\ee
And we have used the generalized Stokes theorem. The remaining contour integrals encircle the cuts 
$\Sigma_a$ running from $\zeta_a$ to $\infty$ (See Fig. \ref{fig:cuts}).

The value of $ \psi_j$ above and below 
\footnote{`Clockwise' and `anticlockwise' may be more precise terms here 
than 'above' and 'below'.
If the cut extend to infinity on the right side then the two terminologies 
are consistent.}
the cut are related by 
\be
{(\psi_j)}_-=\nu_a{ (\psi_j)}_+\quad \nu_a=e^{-2\pi i\Phi_a}
\ee
To see how $\Psi_k$ behaves across the cut $\Sigma_a$ write
\be
\Psi_k(z;\zeta,\xi_0)= \mathbf{\Psi}_{ak} (\zeta)+\Psi_k(z;\zeta,\zeta_a), 
%\quad \mathbf{\Psi}_{ak} (\zeta)= {\Psi}_k (\zeta_a;\zeta,\xi_0)
\ee
where
\be\label{paj}
\mathbf{\Psi}_{ak} (\zeta)= {\Psi}_k (\zeta_a;\zeta,\xi_0)= \int_{\xi_0}^{\zeta_a} d\xi\, {\xi^k }\psi_0(\xi), 
\ee

The first term is a finite\footnote{since  $\Phi_a<1$} constant (independent of $z$)}. 
The second term inherits the $\nu_a$ discontinuity of 
$\psi_j$. It follows that $\Psi_k(z;\zeta,\zeta_a)\bar\psi_j(z)$ 
is continuous across the cut $\Sigma_a$ and does not contribute to the integral in Eq.~(\ref{g}). The metric reduces to
\begin{align}\label{eq:g}
\mbf{g}_{jk}&=\frac i 2 \sum_{a}  \mathbf{\Psi}_{ak} (\zeta) \int_{\Sigma_a} {\bar\psi}_{j} d\bar z
%%  , \quad \mathbf{\Psi}_{ak} (\zeta)= {\Psi}_k (\zeta_a;\zeta,\xi_0)
\\ &= \frac i 2 \sum_{a}  \mathbf{\Psi}_{ak} (\zeta) (1-\bar\nu_a)\big(\bar\Psi_{j}(\infty_a;\zeta,\xi_0)-\bar{\mathbf{\Psi}}_{aj}(\zeta)\big)
\nonumber  \end{align}
 We denote by 
${\Psi}_{j}(\infty_a;\zeta,\xi_0)$ the value attained by ${\Psi}_{j}(z;\zeta;\xi_0)$ as $z$
 tends to infinity in the region between the cuts $\Sigma_a$ and 
$\Sigma_{a+1}$.  {(See Fig. \ref{fig:cuts}.)}
The limit is well defined at infinity provided $j\le D-1$, which is what we need for the metric. 

Rewrite Eq.~(\ref{eq:g}) as a matrix equation
\be\label{eq:g2}
\mbf{g}= \left(  \mathbf{\Psi}_\infty-\mathbf{\Psi}\right)^* \mbf{G}_1  \mathbf{\Psi} , \quad (\mbf{G}_1)_{ab}=
\frac i 2 \delta_{ab}(1-\bar\nu_a)
\ee
 $\mbf{G}_1(\Phi)$ is independent  of $\zeta$.

The $N$-tuples $\mathbf{\Psi}_\infty=\big ({\Psi}_{j}(\infty_{1}),\dots,{\Psi}_{j}(\infty_{N})\big)$
and   $\mathbf{\Psi}=({\Psi}_{1j},\dots,{\Psi}_{Nj})$ are linearly dependent
\be\label{yjf}
{\Psi}_{j}(\infty_{a-1})-{\Psi}_{aj}=\nu_a
({\Psi}_{j}(\infty_{a})-{\Psi}_{aj}),\quad a\in 1,\dots, N
\ee
where  ${\Psi}_{j}(\infty_{N})={\Psi}_{j}(\infty_{0})$.
This comes from 
integrating  ${\psi}_-= \nu_a{\psi}_+$  along $\Sigma_a$.
{Eq.~(\ref{yjf}) too may be written as a matrix equation}
\be\label{yjf2}
\mbf{G}_2 \mbf{\Psi}_\infty=\mbf{\bar G}_1 \mbf{\Psi},\quad (\mbf{G}_2)_{ab}=\frac i 2 (\delta_{ab}\nu_a-\delta_{a,b+1})
\ee 
It follows from Eq.~(\ref{eq:g2}) and Eq.~(\ref{yjf2}) that  the $D\times D$ matrix $\mbf g$ can be factored as
%\margin{check. not manifestly hermitian}
\be\label{gexplicit}
\mathbf{g}(\zeta;\Phi)= \mathbf{\Psi}^* (\zeta;\Phi)\mathbf{G}(\Phi) \mathbf {\Psi}(\zeta;\Phi),\quad\ 
\mathbf{G}= \left(\mathbf{\bar G}_1^*\left(\mathbf{G}_2^{-1}\right)^*-\id\right) \mathbf{G}_1
\ee
where:
\begin{itemize}
\item  $\mathbf{\Psi}(\zeta,\Phi)$ is   an $N\times D$ holomorphic (matrix) function whose matrix elements are given 
{ in Eq.~(\ref{paj}).}
%\be\label{paj}
%\mathbf{\Psi}_{aj}(\zeta)= \Psi_j(\zeta_a;\zeta ,\xi_0) %\quad \Psi_j(z;\zeta,\xi_0)=\int_{\xi_0}^z \frac{x^j dx}{(x-\zeta_1)^{\Phi_1}
%%\dots(x-\zeta_N)^{\Phi_N}}
%\ee

%\item $\mbf{\Psi}_{aj}$ is finite 
\item  $\mathbf{G}(\Phi)$ is an $N\times N$  matrix which is independent of the controls $ \zeta\in\mathbb{C}^N$.

\item Since $\mathbf{g}$ is a $D\times D$ positive matrix, $\mathbf{G}$ must be an hermitian matrix. It must have 
at least $D$ positive eigenvalues and the  image of $\mathbf{\Psi}$ must lie in the ``positive cone" of $\mathbf{G}$.
In fact one may show that $\mathbf{G}$ has exactly $D$  positive eigenvalues.
%%%%%%%%%%%%%%%%%%%%%%%

\item 
The definition of $\mbf{\Psi}_j$ {as a primitive integral in Eq.~(\ref{F}) allows addition of
an arbitrary  integration constant (possibly $j$-dependent) corresponding to a free  choice of $\xi_0$.
Change of this choice will} change the columns of $\mathbf{\Psi}$ by constant columns:
\be\label{psigauge}
\delta\mathbf{\Psi}= \left(
\begin{array}{ccc}
\delta c_1& \dots&\delta c_D\\
\dots&\dots&\dots\\
\delta c_1& \dots&\delta c_D
\end{array}
\right)
\ee
Since changing $\xi_0$ must not affect the metric, it follows that the kernel of $\mathbf{G}$ contains the  vector $(1,\dots,1)^t$. 
It is in fact spanned by it. 
%%%%%%%%%%%%%%%%%%%
\item

One convenient `gauge' choice is $\xi_0=\zeta_N$ which makes the last row of $\mathbf{\Psi}_{aj}(\zeta)$ vanish.
As a result  Eq.~(\ref{gexplicit}) takes the form 
{$g=\check{\mathbf{\Psi}}^*\check{\mathbf{G}} \check{\mathbf {\Psi}}$ }
where $\check{\mathbf {\Psi}}$ is $(N-1)\times D$ and $\check{\mathbf{G}}$ is  $(N-1)\times(N-1)$.
\item
{An explicit expression for $\mathbf{G}$ is} 
\be\label{GGN}
\mbf{G}_{ab}=\frac{\sin\left(\pi\Phi_a \right)}{\sin\left( \pi\Phi_T\right)}\times
\begin{cases}
-\sin\left(\pi(\Phi_T-\Phi_a)\right)& a=b\\
\sin\left(\pi\Phi_b\right)
\exp\left[i\pi\left(\Phi_T-\sum_{c=a}^{b-1}(\Phi_c+\Phi_{c+1})\right)\right]&a<b
\end{cases}
\ee
The values for $N\geq a>b\geq 1$ may be deduced from hermiticity condition $G_{ab}=G_{ba}^*$.
Alternatively the same value may be found from the relation $G_{ab}=G_{a,b+N}$.
%\margin{\red{ factor 2 pi}}

\item
When all the fluxes are identical $\mathbf{G}$ is a T\"oplitz matrix, i.e. constant along the diagonals, 
	\be
	\mathbf{G}_{ab}= G_{a-b}, \quad a,b\in 1,\dots,N
	\ee
Explicitly, if each fluxon carries $\Phi_a=\Phi$, then	for ,$0<k\leq N$,
\be\label{GN}
G_0=-{\sin\left( {\pi\Phi} \right)\sin\left({\pi\Phi (N-1)}\right)\over \sin(\pi N \Phi)},\quad
G_k=	e^{i \pi\Phi(N-2k)}\frac{\sin^2\left({\pi \Phi}\right)}{ \sin(\pi N \Phi)}
\ee
\item 
Away from the threshold for appearance of a new zero-mode, $\Phi_T\in\mathbb{Z}$, the elements of $\mathbf{G}$
are well defined and free of singularities. (As are the elements of $\Psi,\mbf{g}$.)

\end{itemize}

%%%%%%%%%%%%%%%%%%%%%%%%%%%%%%%
\section{Three subcritical fluxes}\label{3flux}
%%%%%%%%%%%%%%%%%%%%%%%%%%%%%%%%%%

	\begin{comment}
	\begin{exa}
	For three half fluxes: $\Phi_1=\Phi_2=\Phi_3=1/2$ at $\zeta_1=0, \ \zeta_2=1,\ 	\zeta_3=u$
	\begin{align}\label{1/2}
	\Psi (z;u;0)&=\int_0^z \frac{dx}{ \sqrt{x (x-1) (x-u)} }\nonumber \\
	&=-2 i K(1-u)+2i F\left(\left. i \sinh^{-1}(1/\sqrt{z-1})\right|1-u\right)
	\end{align}
	\href{http://www.wolframalpha.com/input/?i=Elliptic+K}
	{$K(m)$} and 
	\href{http://www.wolframalpha.com/input/?i=Elliptic+F}
	{ $F(\phi|m)$} are elliptic functions and
	\[
	\sinh^{-1}(1/\sqrt{x-1})= \log (\sqrt x+1)-\frac{i}{2}  \log (x-1)
 	\] 
	\end{exa}
	\end{comment}

%\end{document}

For three fluxons one can find explicit expressions for the metric $\mbf{g}$ and the curvature. 
Exploiting translation rotation and dilatation symmetries allow us to fix the location of two fluxons at will.
We shall therefore assume the three fluxon are located at $\zeta_1=0,\ \zeta_2=1,\  \zeta_3=u$.

%Choose $c=0$.  This makes $\Psi_{1j}=0$. 

\subsection{The abelian case: $1<\Phi_T<2$}
%%%%%%%%%%%%%%

Choosing $\xi_0=\zeta_1=0$ leads in the case $D=1$ to the following
\be\label{ucg}
\mathbf{\Psi}=
\left(
\begin{array}{c}
0  \\
u^{-\Phi_3}{\Gamma\left(1-\Phi_1\right) \Gamma\left(1-\Phi_2\right) \over \Gamma\left(2-\Phi_1-\Phi_2\right)}
{\,}_2\tilde{F}_1\left(1-\Phi_1 ,\Phi_3 ;2-\Phi_1-\Phi_2;{ 1\over u}\right) \\
u^{1-\Phi_1-\Phi_3}{\Gamma\left(1-\Phi_1\right) \Gamma\left(1-\Phi_3\right) \over \Gamma\left(2-\Phi_1-\Phi_3\right)}
{\,}_2\tilde{F}_1\left(1-\Phi_1 ,\Phi_2 ;2-\Phi_1-\Phi_3;u\right)  \\
\end{array}\right)
\ee
{Here} \href{http://functions.wolfram.com/HypergeometricFunctions/Hypergeometric2F1/}{${\,}_2\tilde{F}_1$}
is a hypergeometric function.
For three identical fluxes  $\Phi_a=\Phi$ this reduces into
\be\label{Psi}
\mathbf{\Psi}={\Gamma\left(1-\Phi \right)^2 \over\Gamma\left(2-{2\Phi}\right)}
\left(
\begin{array}{c}
0  \\
u^{-\Phi} F\left(\frac 1 u\right) \\
u^{1-2\Phi}  \,F(u)  \\
\end{array}\right),\quad F\left( u\right)={\,}_2\tilde{F}_1\left(1-\Phi ,\Phi ;2-{2\Phi} ;{ u}\right)
\ee
In particular in the special case $\Phi= 1/2$ it becomes

\be\label{hk}\mathbf{\Psi}=
\left(  \begin{array}{cc}
0  \\
{2\over\sqrt{u}}K\left({1\over u}\right)  \\
2K(u)   \\
\end{array}\right)
\ee
%% with $K$ the an elliptic integral.
with \href{http://www.wolframalpha.com/input/?i=Elliptic+K}{$K(m)$}  the complete elliptic integral of the first kind.
Using Eq.~(\ref{GN})  for $\mbf{G}$, one then finds a simple formula for the metric
\be\label{fig:r}
\mbf{g}(u)=8 Re\left( K(u)K(1-u)^*\right)
\ee
The associated curvature is plotted in Fig.~(\ref{fig:curvature}).

Since $\mathbf{\Psi}$ is defined only up to addition of an arbitrary ($u$-dependent) multiple of $\mbf{1}_N$,
one may write down various alternative expressions to Eq.~(\ref{ucg}). Using the following
 \begin{align}
\mathbf{\Psi}=&
{\pi\Gamma(\Phi_1+\Phi_3-1)\over\Gamma(\Phi_1)\Gamma(\Phi_3)}
\left(\begin{array}{c}
{e^{i\Phi_1\pi}\over\sin(\Phi_1\pi)} \\ 0 \\ -{e^{-i\Phi_3\pi}\over\sin(\Phi_3\pi)} \\
\end{array}\right)
u^{1-\Phi_1-\Phi_3}\,_2\tilde{F}_1\left(1-\Phi_1,\Phi_2,2-\Phi_1-\Phi_3;u\right)
 \nonumber \\
+&{\Gamma(1-\Phi_2)\Gamma(1-\Phi_1-\Phi_3)\over\Gamma(2-\Phi)}e^{-i\pi \Phi_3}
\left(\begin{array}{c}
0 \\ 1 \\ 0 \\
\end{array}\right)
\,_2\tilde{F}_1\left(\Phi_3,\Phi-1,\Phi_1+\Phi_3;u\right)
 \nonumber \end{align}
with $\mbf{G}$  the $3\times 3$ matrix given in Eq.~(\ref{GGN}), leads to expressing
the $1\times 1$ metric $\mbf{g}$ as a combination of two squares
\begin{align}\mbf{g}=&
\pi{\Gamma(1-\Phi_2)\over\Gamma(\Phi_2)}
{\Gamma(1-\Phi_1-\Phi_3)\over\Gamma(\Phi_1+\Phi_3)}
{\Gamma(\Phi-1)\over\Gamma(2-\Phi)}
\left|\,_2\tilde{F}_1\left(\Phi_3,\Phi-1,\Phi_1+\Phi_3,u\right)\right|^2
\nonumber \\
&+\pi{\Gamma(1-\Phi_1)\over\Gamma(\Phi_1)}
{\Gamma(1-\Phi_3)\over\Gamma(\Phi_3)}
{\Gamma(\Phi_1+\Phi_3-1)\over\Gamma(2-\Phi_1-\Phi_3)}
\left|u^{1-\Phi_1-\Phi_3}\,_2\tilde{F}_1\left(1-\Phi_1,\Phi_2,2-\Phi_1-\Phi_3,u\right)\right|^2
\nonumber \end{align}
For $1<\Phi_T<3$ this is always positive.

%%%%%%%%%%%%%%
\color{black}
\subsection{The non-abelian case: $2<\Phi_T<3$}
One can also get explicit formulas for the non-abelian case.
%For $\Phi_T>2$ we have $D=2$ and 
$\mathbf{\Psi}$ is $3\times2$:
\begin{comment}
\be\label{Psig}
\mathbf{\Psi}=u^{-\Phi_3}\left( \begin{array}{cc}
0 & 0 \\
{\Gamma(1-\Phi_1)\Gamma(1-\Phi_2)\over\Gamma(2-\Phi_1-\Phi_2)}\,_2\tilde{F}_1\left(1-\Phi_1,\Phi_3,2-\Phi_1-\Phi_2,{1\over u}\right) & 
{\Gamma(2-\Phi_1)\Gamma(1-\Phi_2)\over\Gamma(3-\Phi_1-\Phi_2)}\,_2\tilde{F}_1\left(2-\Phi_1,\Phi_3,3-\Phi_1-\Phi_2,{1\over u}\right) \\
u^{1-\Phi_1}{\Gamma(1-\Phi_1)\Gamma(1-\Phi_3)\over\Gamma(2-\Phi_1-\Phi_3)}\,_2\tilde{F}_1\left(1-\Phi_1,\Phi_2,2-\Phi_1-\Phi_3,u\right)   &  
u^{2-\Phi_1}{\Gamma(2-\Phi_1)\Gamma(1-\Phi_3)\over\Gamma(3-\Phi_1-\Phi_3)}\,_2\tilde{F}_1\left(2-\Phi_1,\Phi_2,3-\Phi_1-\Phi_3,u\right)    \\
\end{array}\right) \ee
%%%%%%%
\end{comment}
\be\label{Psig}
\mathbf{\Psi}=u^{-\Phi_3}\left( \begin{array}{cc}
0 & 0 \\
\frac{\Gamma(1-\Phi_1)\Gamma(1-\Phi_2)}{\Gamma(2-\Phi_1-\Phi_2)}\,A & 
\frac{\Gamma(2-\Phi_1)\Gamma(1-\Phi_2)}{\Gamma(3-\Phi_1-\Phi_2)}\,B\\
u^{1-\Phi_1}\frac{\Gamma(1-\Phi_1)\Gamma(1-\Phi_3)}{\Gamma(2-\Phi_1-\Phi_3)}\,C   &  
u^{2-\Phi_1}\frac{\Gamma(2-\Phi_1)\Gamma(1-\Phi_3)}{\Gamma(3-\Phi_1-\Phi_3)}\,D  \\
\end{array}\right) \ee

where
\begin{align*}
A={\,}_2\tilde{F}_1\left(1-\Phi_1,\Phi_3,2-\Phi_1-\Phi_2,\frac 1 u\right) &\quad
 B={\,}_2\tilde{F}_1\left(2-\Phi_1,\Phi_3,3-\Phi_1-\Phi_2,\frac 1 u\right) \\
C={\,}_2\tilde{F}_1\left(1-\Phi_1,\Phi_2,2-\Phi_1-\Phi_3,u\right)&\quad 
D={\,}_2\tilde{F}_1\left(2-\Phi_1,\Phi_2,3-\Phi_1-\Phi_3,u\right)  
\end{align*}
%%%%%%%%%%%%%%%%%%%%%%%%%%%%%%
When all three  fluxes are identical $\Phi_a=\Phi$ this becomes
%\margin{funny factors 2}
%%%%%%%%%%%%%%%%%%%%%%%%%%%%%%%
\be\label{Psi2}
\mathbf{\Psi}={\Gamma\left(1-\Phi \right)^2\over2\Gamma\left(2-{2\Phi}\right)}
\left(
\begin{array}{cc}
0 & 0 \\
 2u^{-\Phi}F\left(\frac 1 u\right) & \,u^{-\Phi}G\left(\frac 1 u\right) \\
2u^{1-2\Phi}  \,F(u) &   u^{2-2\Phi}  \,G(u) \\
\end{array}\right)
\ee
where 
\[
F\left( u\right)={\,}_2\tilde{F}_1\left(1-\Phi ,\Phi ;2-2\Phi;{ u}\right),\quad 
G(u)={\,}_2\tilde{F}_1\left(2-\Phi ,\Phi ;3-2\Phi;u\right)
\]
$\mbf{G}$ is again $3\times 3$ given by Eq.~(\ref{GGN}) and $\mbf{g}$ is $2\times 2$.

%%%%%%%%%%%%

%%%%%%%%%%%%%%%%%%%%%%%%%%%%%%%%%%%%
\section{An abstract construction of the adiabatic connection.}

{In this appendix we give another (more abstract) construction of the connection described in { Sec. \ref{sec:connection1} 
and Sec. \ref{sec:square}} corresponding to the free  zero modes around point-like fluxons.
In particular it shows that in general one may embed our $D_f$-dimensional bundle into a flat $N-1$ bundle.}

Let $V_0$ be the fixed $(N-1)$-dimensional complex vector space $\mathbb{C}^N/\mathbb{C} \,\mbf{1}_N$ where 
$\mbf{1}_N=(1,1,...1)^t$.
Since $ker(G)=\mathbb{C}\,\mbf{1}_N$, the $N\times N$ hermitian matrix $G$ defines a pseudo (hermitian) metric on $V_0$.
Let ${\cal M}=\{(\zeta_1,\zeta_2,...\zeta_N)\in\mathbb{C}^N|\; \forall a\neq b\;\; \zeta_a\neq\zeta_b\}$ be the space of 
possible positions
of $N$ fluxons.  Consider the trivial bundle $E_0={\cal M}\times V_0$.
%%%%%%%%%%%%%%
%%%%%%%%%%%
For each $j$ the vector function $\Psi_j(\zeta)=(\Psi_{1j},\Psi_{2j},...\Psi_{Nj})$ defines a (multivalued) section of $E_0$. We shall 
denote this section 
by $\Psi_j$ as well although it is actually an equivalence class under quotienting by $\mbf{1}_N=(1,1...1)^t$.

At each point $\zeta\in{\cal M}$ the vectors $\Psi_0(\zeta),...\Psi_{D-1}(\zeta)$ generate a $D$-dimensional subspace
$V_\zeta$ of $V_0$. These spaces make up together a $D$-dimensional sub-bundle $E$ of $E_0$.
The restriction of $G$ to $E$ is a positive definite hermitian metric. This follows from the fact that
$g_{ij}=\Psi_i^*G\Psi_j$ is the hilbert space metric on our Pauli zero-modes.
In particular it follows that $E^\perp$ the $G$-orthogonal complement of $E$ is well defined  and hence
also the $G$-orthogonal projection $Q:E_0\rightarrow E$.
In fact one may write explicitly $Q=\sum g^{ij}(\Psi_i\otimes\Psi_j^\dag) G$ where $g^{ij}$ is the inverse of
the matrix $g_{ij}=\Psi_i^\dag G \Psi_j $.

%%%%%%%%%%%%%%%%%%%%%
%%   The fact that $Q$ is $G$-orthogonal imply that $Q^\dag=GQG^{-1}$.
%%   The fact that the sections  $\Psi_1(\zeta),...\Psi_D(\zeta)$  which generate $E$ are holomorphic imply
%%%
%%%
%%   that $Q(\bar{\partial}Q)=\bar{\partial}Q$. From which follows also $(\partial Q)Q=\partial Q.$
%%   (Interpreting $\partial,\bar{\partial},Q$ as operators, one may rewrite this as $Q\partial Q=Q\partial$..)
%%%%%%%%%%%%%%%%%%%%%%%%

As $E_0$ is trivial it is natural to use the trivial connection ${\cal D}_0=d$ on it.
The projection $Q:E_0\rightarrow E$ then defines a connection ${\cal D}=Qd$ on $E$.
Consider a general section $\Psi=\sum p_k\Psi_k$ of $E$.
Using the fact that $d\Psi_k=\partial\Psi_k$ we find that
the covariant derivative is given by:
$${\cal D}\Psi=Q\; d\Psi=\sum g^{ij}\Psi_i\otimes\Psi_j^\dag G
(\Psi_k dp_k+p_k \partial\Psi_k)=
g^{ij}\Psi_i\left( (dp_k)+p_k\partial\right)g_{jk}$$
The equation $\mathbf{v}\cdot{\cal D}\Psi=0$ for parallel transport thus takes the form
$$\mathbf{v}\cdot\left( g dp+(\partial g)p\right)=0$$
which is exactly identical to the transport equation Eq.~(\ref{yr}).

%%%%%%%%%%%%%%%%%%%%%%%%%%%%%%%%%%%

%%%%%%%%%%%%%%%%%%%%%%%%%%%%

\section{The holonomy of a rotating fluxon-An intriguing factor}\label{appendix:rot}

\begin{figure}[h]
  \centering
\includegraphics[width=6 cm]{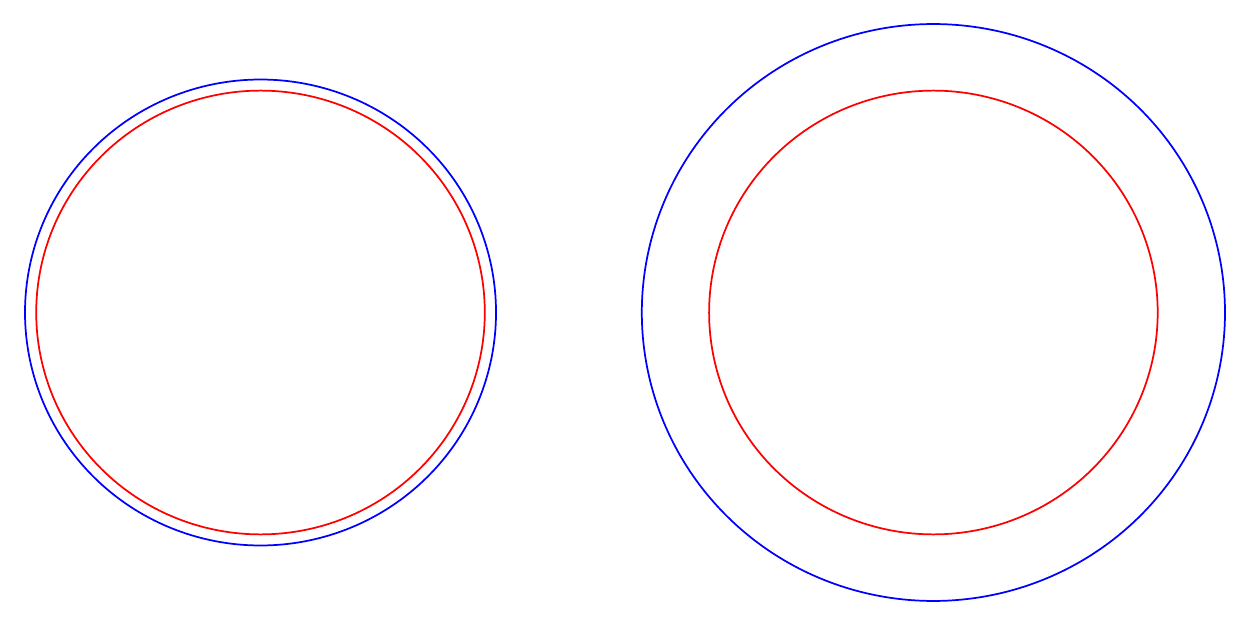}
 \caption{The fluxon is modelled as two counter-circulating charged cylinders of radius  $R$ and charge density $\pm\sigma$ . 
The red cylinder rotates clockwise and the blue counter-clockwise with the same angular velocity. Rotating the fluxon causes the red 
cylinder to rotate faster and the blue cylinder slower. This creates a voltage difference between the inside 
and outside of the fluxon.   }\label{box}
\end{figure}

Consider adiabatically turning one of the flux tubes around itself once. 
To find the holonomy of zero energy bound states we first 
need to find the  electric and magnetic fields  generated by adiabatic  
rotation at angular rate $\delta\Omega$. To find these, we need a model of a fluxon.   
Consider the following simple model\footnote{We do not claim universality and the results may be model dependent.} 
of  fluxon, shown in  Fig.~\ref{box}: Two concentric thin  cylinders of radius $R$ with charge $\pm Q$ (per unit length),  
and charge density $\sigma=\pm Q/(2\pi R)$, rotating at constant angular velocity $\pm \omega$.

Since the overall charge vanishes and the fields are time independent, there is no electric field. The magnetic field is 
static and it satisfies, (Recall $c=1$)
\[
\nabla\cdot B=0, \quad \nabla\times B =4\pi j =8\pi\sigma \omega R \delta(|\mathbf{x}|-R)\hat \theta
\]
{leading to a jump} in the boundary conditions
\[
B(R_-)-B(R_+)={8\pi \sigma \omega R}=4Q\omega, \quad
\]
 Assuming $B(\infty)=0$ we then have
\be\label{flux}
B(\mbf{x})=\begin{cases}  { 4 Q \omega }&|\mbf{x}|<R\\
0&|\mbf{x}|>R\end{cases}
\ee
It follows that the flux, per Eq.~(\ref{zero-modes}), is
\be
\Phi_T =2 e Q R^2 \omega 
\ee
Consider what happens  when one adiabatically rotates the whole arrangement by $\delta \Omega$ so the two cylinders rotate 
at different angular velocities.

To figure out the addition  of angular velocities $\omega$  and $\pm\delta\Omega$, let
\[
\gamma=\frac 1 {\sqrt{1-(\omega R)^2}}, \quad \omega R<1
\]
If $\gamma\approx 1$ then addition gives $\omega'=\omega\pm\delta\Omega$.  
But if we allow $\gamma\gg 1$, the rule follows from additivity of the rapidity  $\tanh\theta=v$. 
One finds (assuming $\delta\Omega$ small)
\[
\delta\omega=\omega'-\omega =\pm \frac {\delta\Omega}{\gamma^2}
\]
If this was all that happened, rotating the fluxon would have no effect on the fields (to order $\delta\Omega$). However, this is not all. 
Relativistic Lorentz contraction implies that the geometry  of the cylinders must change\footnote{Rigid bodies are inconsistent with 
special relativity.}.
The perimeter of the cylinders should contract by the usual rule.
As the embedding space remains Euclidean the radius needs to  adjust to accommodate the contraction.
For a cylinder of finite width this would inevitably lead to nontrivial internal stresses, but in the zero width limit
we consider here this issue can be ignored.
Thus  if $R$ denotes the radius for the cylinder rotating with $\omega$ then the contraction of the radii is given by 
\[
R' \gamma(\omega' R')= R \gamma(\omega R)
\]
It follows that, to first order in $\delta\Omega$
\[
R' =R\left(1-\frac{\delta\gamma}{\gamma}\right)=
R\left(1- \gamma^2{R^2\omega}\delta\omega \right)=
R\left(1\mp \delta\Omega {R^2\omega} \right)
\]
Hence
\[
\delta R= \mp \delta\Omega {R^3\omega} 
\]
This imply that in the annulus between the two cylinders there is a radial  electric field and hence a potential difference between the 
inside and the outside of the fluxon :
\be\label{2}
 \delta V= 2 Q \log \frac{R+\delta R}{R-\delta R}\approx 4 Q\frac {\delta R} R=4Q {R^2 \omega} \delta\Omega =
\frac 2 e\Phi_T \delta\Omega
\ee
Integrating over the time needed to complete one full rotation gives
\[
\int e\delta V dt=  {4\pi} \Phi_T 
\]
%%%%%%%%%%%%%%%%HHHHHHHHHHHHHH%%%%%%%%%%%%
Consider a charged particle having wave function $\psi$ in the presence of the fluxon.
The above suggests that fluxon rotation would induce a phase on the part of the wave function which is inside the fluxon.  
If the evolution is adiabatic this relative phase can be translated into the overall phase  

\[\frac  {4\pi q}e \Phi_T ,\quad q=e\int_{r<R}|\psi|^2d^2r
\]
 $q$ is the (fraction) of charge inside the fluxon. 
The phase depends on the total charge inside the fluxon but is independent of how it is  distributed there. 

It is instructive to contrast this result with what one expects  from the  Aharonov-Bohm effect.  A (classical, localized) magnetic flux 
encircling a localized (quantum) charge $q$  gives half this phases. 
Not only is the factor 2 intriguing but, even more importantly, the Aharonov Bohm argument implies that  the phase 
should depend also on the distribution of charge inside the fluxons.

%%  \deleted{  we found here depends only on the total amount of charge present inside the radius $R$, while for a magnetic 
%%  flux spread uniformly in the disc of radius $R$ one expects the A-B phase to depend continuously on the distance of the charge 
%%  from the disc center.}
%%% on the whole distribution of charge inside the disc.}%%

%%%%   \mathbf{v}\times\mathbf{B}

\section{Moving the flux along a general vector field}\label{sec:d}

In the discussion of moving fluxons,  Section \ref{7tuygd},  it was assumed for simplicity that $\mbf{v}$ stands for a vector rather 
than a vector field.
As a result {self} rotations of fluxons were not permitted\footnote{{In the discussion of section\ref{ukhd}
only the position $\zeta_a$ of the fluxons centers were rotated.}}. It is in fact possible to generalize {parts of our} arguments
to arbitrary vector fields. The aim of this appendix is to explain this. It is worthwhile to note that the following
does not even require the introduction of a Riemann metric as the argument are completely independent of it.

By a slowly moving fluxon we shall mean a fluxon whose  magnetic field $B$ is being 'dragged' along the vector field $\mathbf{v}$.  
More precisely, this is described mathematically by writing
$$\dot{\mathbf{B}}=-{\cal L}_{\mathbf{v}} \mathbf{B}$$
where ${\cal L}_{\mathbf{v}}$ stands for the Lie derivative along $\mathbf{v}$.
In two dimensions $B$ is a scalar density and using explicit form of the Lie derivative gives
$$-\dot{B}=(\mathbf{v}\cdot\nabla)B+B(\nabla\cdot \mathbf{v}).$$
The first term describes moving along $\mathbf{v}$ while the second makes $B$ behave as a density in cases where the 
flow defined by $\mathbf{v}$ does not preserve volume. In particular this relation guarantees that the total flux is unchanged.
In case of translations or rotations the second term vanishes anyway.

The corresponding vector potential $\mathbf{A}$ is of course not determined uniquely, but it is most convenient 
to assume it is dragged in a similar way. This leads to
$$\dot{\mathbf{A}}=-{\cal L}_{\mathbf{v}} \mathbf{A}$$
Since $\mathbf{A}$ is a {covariant} vector the explicit form in vector notation is 
$$-\dot{A}_\mu=v^\lambda\partial_\lambda A_\mu+A_\lambda\partial_\mu v^\lambda.$$
The second term represents the required rotation of the vectorial components of $\mathbf{A}$.
This relation may also be expressed as 
$$\dot{A}_\mu=-\partial_\mu(v^\lambda A_\lambda)+v^\lambda B_{\lambda\mu}$$
Substituting this into Eq.(\ref{fields}) for the electric field we find
$$E_\mu=\partial_\mu\left(A_0+v^\lambda A_\lambda\right)+v^\lambda B_{\lambda\mu}$$
As in section \ref{7tuygd} it follows that the relation  $\mathbf{E}=-\mathbf{v}\times \mathbf{B}$ 
is consistent with the choice $A_0=-\mathbf{v}\cdot \mathbf{A}$.
This holds generally regardless of whether $\mathbf{v}$ is a rigid motion or a deformation, and of whether it is constant or
time dependent. It does not even matter here whether space is flat or curved. 
(This may however matter when one considers the spinor $\psi$.)

We remark that when considering a closed loop in deformation space, the above construction guarantees that
the potential $(\mathbf{A},A_0)$ also complete a closed loop (rather then only the fields $\mathbf{B},\mathbf{E}$).
\begin{rem}
In the complex notation $A={1\over2}(A_1-iA_2),v=v^{1}+iv^{2}$ the result takes the form
%%  $$-\dot{A}=(v\partial+\bar{v}\bar{\partial})A+A\partial v+\bar{A}\partial\bar{v}$$
$$\dot{A}={i\over2}B\bar{v}-\partial(Av+\bar{A}\bar{v})$$
\end{rem}

%%%%%%%%%%%%%%%%%%%%%%%%%%%%%%%%%%%%%%

%%%%%%%%%%%%%%%%%%%%%%%%%%%%%%%%%
%\bibliography{fluxons}{}
%\bibliographystyle{plain}
%%%%%%%%%%%%%%%%%%%%%%%%%%%%%%

\end{document}